\documentclass[11pt]{article}
\usepackage[utf8]{inputenc}
\usepackage[english]{babel}
\usepackage{amssymb}
\usepackage{mathrsfs}
\usepackage{stmaryrd}
\usepackage{tabularx}
\usepackage{booktabs}
\usepackage{times}
\usepackage{setspace}
\usepackage{authblk}
\usepackage{placeins}
\usepackage{xfrac}
\usepackage{graphicx}
\usepackage{float}
\usepackage[T1]{fontenc}
\usepackage[final]{pdfpages}
\usepackage{pdflscape}
\usepackage[a4paper,top=3cm,bottom=2cm,left=3cm,right=3cm,marginparwidth=1.75cm]{geometry}
\usepackage{amsmath, amssymb}
\usepackage{amsthm}
\usepackage[colorinlistoftodos]{todonotes}
\usepackage[colorlinks=true, allcolors=black]{hyperref}

\usepackage[tight]{shorttoc}

\usepackage{hyperref}
\usepackage{mathabx}
\usepackage[bottom]{footmisc}
\usepackage{eurosym}
\usepackage{tocloft}
\newcommand{\R}{\mathbb{R}}

\newcommand{\N}{\mathbb{N}}
\newcommand{\E}{\mathbb{E}}
\newcommand{\V}{\mathbb{V}}
\newcommand{\p}{\mathbb{P}}

\newcommand{\cov}{\operatorname{Cov}}

\newcommand{\cP}{{\mathcal P}}

\newcommand{\e}{\varepsilon}

\newcommand{\cB}{{\mathcal B}}

\newcommand{\cM}{{\mathcal M}}

\newcommand{\cC}{{\mathcal C}}

\newcommand{\cL}{{\mathcal L}}
\newcommand{\cR}{{\mathcal R}}

\newcommand{\erv}{\operatorname{ERV}}

\newcommand{\Exp}{\operatorname{Exp}}
\newcommand{\Par}{\operatorname{Par}}
\newcommand{\ki}{\kappa_I}
\newcommand{\kd}{\kappa_D}
\newcommand{\ks}{\kappa_S}
\newcommand{\rhm}{\rho_{max}}

\everymath{\displaystyle}

\theoremstyle{definition}

\newtheorem{defi}{Definition}[section]
\newtheorem{lem}{Lemma}[section]
\newtheorem{prop}{Proposition}[section]
\newtheorem{cor}{Corollary}[section]
\newtheorem{rem}{Remark}[section]

\usepackage{hyperref}
\hypersetup{	
colorlinks=true,
breaklinks=true,
urlcolor= purple,
linkcolor= black,
citecolor= brown,
}

\usepackage{lipsum}

\title{An INAR(1) Risk Model with by-claims applied to the PFAS issue}

\author[1,2]{Charles Minier\thanks{charles.minier41@gmail.com (corresponding author)}}
\author[3]{Stéphane Loisel \thanks{stephane.loisel@lecnam.net}}
\author[4]{Naushad Mamode Khan\thanks{n.mamodekhan@uom.ac.mu}}

\affil[1]{Université Claude Bernard Lyon 1, Laboratoire de Sciences Actuarielle et Financière, Institut de Science Financière et d’Assurances (ISFA), Lyon, France}
\affil[2]{Chaire DIALog, Fondation du Risque, Institut Louis Bachelier, Paris, France}
\affil[3]{Conservatoire National des Arts et Métiers (CNAM), Paris, France}
\affil[4]{Department of Economics and Statistics, University of Mauritius, Moka, Mauritius}

\begin{document}

\maketitle

\begin{abstract}
We propose there a new type of unidimensional discrete-time risk model in ruin theory involving both main claims and by-claims, where the latter ones are counted thanks to an INAR(1) process. In this way, the model embodies a contagion phenomenon, which may be used to describe the damages caused and covered by an insurance company ensuing the spreading of PFAS compounds in the environment.
We present mostly asymptotic ruin probabilities, and multiple comparisons with other models derived from the aforementioned one, in particular with two models where the dependence assumptions are relaxed.

\smallskip

\end{abstract}

\vspace{3em}


\newpage
\tableofcontents

\newpage
\section{Introduction}

Motivated by emerging risk from the perfluorinated compounds (commonly named PFAS), and for the purpose of doing a first work on this issue in 
ruin theory, we propose there an innovative discrete-time risk model, involving both main claims and by-claims, the latter ones counted thanks to an INAR(1) (integer-valued autoregressive) time series.

Due to the speed at which the damages may spread widely from emissions of these chemical products in the environment, it is an important question that encourages the settlement of some specific insurance policies nowadays.
In fact, the current importance of the PFAS issue does not lie in the age of these products or the knowledge of their hazards, but rather in the solutions tried to deal with such pollution and confine their harmful effects.
Relying only on the data available at the moment, it is difficult to predict the related damages which will be revealed, even in the few subsequent years, but the scale of contamination remains surely underestimated, according to many specialists (see \cite{labadie2020contaminants} or \cite{cartedecontaminationanimauxEWG} for instance).

Moreover, in actuarial science, contagion models have been used to describe phenomena which generate risk propagation, or contamination, and thus may cause clusters of claims.
We can cite, for instance, cyber risks, which may be represented thanks to Hawkes processes (see e.g. \cite{bessy2021multivariate}, \cite{hillairet2023expansion}), or epidemiological risks (see \cite{lefevre2017epidemic}, \cite{barigou2020parsimonious}, or \cite{lefevre2022risk}). 
In this paper, we explain how our model, structured thanks to an INAR(1) process, may also describe a contagion phenomenon.
\bigskip

In order to represent such a risk propagation, it is possible to refine the original model of Cramér-Lundberg by adding a second kind of loss, the by-claims, which always arise from a main claim when they occur.
Indeed, for an insurance company, the events which cause serious damages and great amounts to reimburse engender generally two types of claims: the "main claims" which are settled quickly, and the so-called "by-claims" which often correspond to delayed claims.
Such by-claims may describe allocated loss adjustment expenses (ALAE), or embody cost revising due to further events. An example of revising may be found typically in motor insurance, where the claims concern mainly material and health damages: the costs for the first ones are usually settled in due time, while the bodily injury claim can be delayed, because of the time needed to provide the necessary medical care and evaluate the total amount spent.

Some first models with delayed or by-claims were studied 
by \cite{waters1985ruin}, \cite{yuen2001ruin}, or \cite{yuen2005ultimate}. The first one studies only the idea of delay, whereas the last two introduce both secondary claims and delays.
While, other research articles (\cite{macci2006large}, \cite{zou2013probability}, \cite{li2013pairwise}, \cite{fu2016asymptotic}, \cite{li2017note}, \cite{yang2019asymptotic}, \cite{zhang2021asymptotics}, \cite{li2023asymptotic}) discuss about the by-claims in continuous-time models. 

However, most of these articles assume that each main claim engenders exactly one by-claim. 
The only exception is \cite{li2023asymptotic}, which proposes a random number of by-claims coming from each main claim, but the variables counting the by-claims are assumed to be independent. 
This assumption would fit if we only considered ALAE for instance, but in our context, it is limiting: since the perfluorinated compounds are considered "forever pollutants" and may spread fast through the waters or the food chains, it is likely that one source event of toxic emission will induce many potential collateral damages. 
\bigskip

Using an INAR(1) process incorporates this possibility of contagion with a dependence structure. This type of process was introduced for the first time in the work of \cite{al1987first}.
Actually, in the existing ruin theory literature, researchers have proposed an ample class of discrete-time risk models where the number of settled claims is described thanks to time series and a thinning-based dependence structure
(see e.g. \cite{Cossette2010DiscreteTimeRM},  \cite{Shi2014AnAM}, \cite{ma2015bidimensional}, \cite{hu2018risk}, \cite{guan2022analysis}). 
However, contrary to the work presented in this article, all of them include only one type of claim. 
Moreover, contrary to \cite{Cossette2010DiscreteTimeRM} or \cite{Shi2014AnAM} for instance, which use an INAR structure, we attempted to build a non-stationary process, since we try to represent a new category of risk which is still difficult to apprehend.
\smallskip

Thus, our model aims at describing the concrete behaviour of an insurance company willing to handle the PFAS issue, by taking into consideration the emissions and their consequences, and the related liabilities.
Furthermore, we try to understand what happens if we reduce one aspect or another of the whole contamination process, notably through a better containment of the potential emissions, the detection of the chemicals, or recovery campaigns.
 
Nevertheless, as a disclaimer, we have to acknowledge we were motivated by this PFAS thematic, but here, the aim of our paper is rather to alert about this issue, and initiate some works in actuarial science about it, while displaying a model which emphasises the different possible levels of risks. Indeed, for the moment, we lack data to complete and apply our theoretical approach. It is mostly due to the fact that is still a new subject for insurance companies, and regulatory solutions are still harshly discussed (in particular, as specified later, some insurers have settled wide exclusions for the moment).
\bigskip

We provide first some properties linked with the net profit condition of the model, and covariances of the counting processes. Then, we derive the existence of an adjustment coefficient when the losses are light-tailed, and it is possible to compute it precisely when these losses are exponential.
We also propose some asymptotic ruin probabilities in the presence of heavy-tailed distributions, and we linked these results with similar ones from models involving by-claims or an INAR process
We also highlight the fact that neglecting the level of dependence should lead to an underestimation of the risk incurred, through the comparison with two other models coming from the main one, where the dependence assumptions are "softened" (especially in terms of adjustment coefficients). 
In addition, we yield some numerical results of ruin probabilities, and through that we emphasize the hazard of understating the emissions or the contamination.
Eventually, we consider other ways to link the concrete issue of the PFAS with our model by slightly modifying it.

The paper is organised as follows: in Section \ref{section de motivation sur les PFAS}, we present some aspects of the chemical compounds, and their link with our study.
In Section \ref{section de présentation du modèle initial}, we present the main risk model of this article. In Section \ref{section avec les premières propriétés}, we give some first properties of this model. In Section \ref{section ruin prob queues fines}, we present some results with light-tailed distributions for the main model.
In Section \ref{section de comparaison des modèles indépendants}, we introduce two other models derived from the main one, where the dependence assumptions are softened, and we compare these three models. 
In Section \ref{section de résultats de ruine avec queue lourde}, we present results related with heavy-tailed distributions, while continuing to compare the three models.
In Section \ref{section résultats numériques article 1D}, we provide numerical results.
Eventually, in Section \ref{section faux INAR(2)} and \ref{section d'ouverture sur le modèle d'incertitude}, we explain how the initial model could be refined to correspond in other ways to such an insurance context.

\section{Some elements about the PFAS compounds}
\label{section de motivation sur les PFAS}

\subsection{Some elements about the PFAS compounds}

The damages related to PFAS molecules (or "Per- and polyfluoroalkyl substances", or "perfluorinated compounds") constitute a burning issue currently.

Indeed, these molecules are extremely stable, and the bond between fluorine and carbon is among the strongest in organic chemistry. Therefore, it allows to build materials which are waterproof or resistant to extreme temperatures for example. However, all these good properties make these molecules almost indestructible: this is why they are called "forever pollutants". In fact, the bigger perfluorinated molecules may decompose in the environment, but always in smaller ones, and these smaller elements will remain persistent (see e.g. \cite{lohmann2020fluoropolymers}). Nowadays, the best ways to clean the environment from these compounds are to collect them, and then stocking them, or using safe disposal methods (thanks to incinerators for example) to destroy or mineralise them, while avoiding re-emissions. However, it remains relatively slow processes regarding the widespread aspect of these products (see e.g. \cite{labadie2020contaminants}).

These molecules are derived from chemical synthesis, namely, they are not extracted from the environment. Manufactured products using PFAS have been marketed for the first time in the 1950s by the American company DuPont de Nemours, and have been used in numerous common objects such as pans, waterproof packaging, or clothes protecting from extreme temperatures.

We may reasonably assume that these forever chemicals have already spread on the whole world's surface, especially through the atmosphere or the waters: they may be found for example in the liver of some animals living in the Arctic region, even if there is no human industrial activity nearby.
\medskip

As mentioned, the present importance of this question is not due to the time since they exist or since we know their hazards (above all for the human health, such as cancers), but rather in the regulatory or practical solutions which are tried in order to limit their damaging effects: detection campaigns, research and implementation of cleaning processes towards the environment, and in the field of insurance companies, specific coverages.

Indeed, even if they exist since the mid-XX$^{th}$ century, it seems that only during the last years some insurance companies have taken into consideration all the high risks coming from PFAS, so as to propose specific coverage (as underlined by Guy Carpenter for instance, in \cite{conferenceGuyCarpenter}). Although pollution policies already exist, the rapid spread and the harmful consequences have encouraged the establishment of policies focused on these perfluorinated compounds. However, for the moment, the exclusions are quite strict. For example, in August 2022, the Lloyd's Market Association published two clauses to exclude almost entirely the responsibility of the (re)insurer for damages related to perfluorinated substances. In June 2024, MunichRe proposed many exclusions, such as for the PFAS manufacturers or public or private entities responsible for water distribution or treatment. Comparatively, in France, the insurance law stipulates that an exclusion has to be limited (\cite{conferenceGuyCarpenter}).

Moreover, in the United States, the coverages proposed to industrialists depend on the period during which their activity occurred, regarding the legislation enacted since 1950 towards the PFAS. They might also be contentions depending on the state where the emissions took place, the state where the damages are signalled, and even the state where the insurance contract was signed (\cite{chesler2021pfas}). Thus, some industrial companies that incorporated perfluorinated into their products and thought to be exempt from this issue could be ensnared this way, according to some reports from GenRe for example (\cite{articleGenRe}). 

In terms of regulation, more and more PFAS species are subject to European or American exclusions, and it should continue to be toughened in the coming years. Yet, the thresholds still look heterogeneous on a global scale.

For instance, in the European Union, some new thresholds of drinking water will be applied in 2026, whereas for food, the current ones have existed since 2023. 
Actually, depending on some specific molecules, and the possible uses made (as a whole, or related to these particular ranges), there could be different levels of ban or acceptance.

Regarding detection, multiple campaigns are led to identify at different scales the presence of molecules in the water or other substances, and sometimes, maps are consequently built to have a better overview, depicting the presence of the most common perfluorinated species in the waters or the bodies of animals, with their concentration at each point measured (\cite{munoz2015spatial}). For instance, a worldwide map is available on the website of the Environmental Working Group (\cite{cartedecontaminationanimauxEWG}).
\medskip

Eventually, the scientific community has to continue to acquire data and specific knowledge about the issue. In particular, it would be good to find more efficient cleaning methods than the ones which are already known.
Moreover, the industrial companies who were producing or using perfluorinated compounds should reduce their usage as much as possible, and change their business model if necessary. However, it seems difficult to prohibit these products completely at the present time for several reasons: in particular, some materials have no reliable alternatives for the moment. Besides, it is possible to educate the consumer about their choices: some brands manufacture products without PFAS similar to some others using these compounds for example, and advertise about it.

\subsection{The possible meanings of the following model}

The model written below in Section \ref{section de présentation du modèle initial} aims at representing the mechanisms and the consequences of the pollution related to the perfluorinated compounds which could be covered by an insurance policy. 

Firstly, a period should correspond to a year (or possibly a month, or a trimester).

The aggregate amount $\sum\limits_{i=1}^{N_k} X_{k,i}$ corresponds to the financial costs linked with civil liability about the pollution caused by some emissions or damages suffered by individuals coming from this pollution, which have been recorded during the $k^{th}$ period.
We can see $N_k$ as the number of such clients targeted by a financial compensation.

On the other hand, the aggregate amount $\sum\limits_{i=1}^{M_k} Y_{k,i}$ corresponds to the gathering of financial costs both to reduce the pollution in the environment, and to manage the health damages caused by emissions caused by a given client during or before the $k^{th}$ period. As mentioned before, the aim of the depollution is to remove as many molecules as possible from the environment, and this could take a long time since the perfluorinated compounds are able to spread widely through the air or the waters. Besides, we can state the related health care concern humans only, since taking care of contaminated animals or plant species would rather correspond to cleaning the environment.

As a remark, both types of claims should likely refer to the whole activity of the client. Indeed, it seems difficult to detect the effects of a single emission, and industrials threatened by financial compensation for their polluting activity should be judged for the whole damages, and not only small parts of that. However, it could also correspond to periods of several years, depending on the evolution of the regulations, according to what may happen in the United States for example (\cite{chesler2021pfas}).
\bigskip

In Equation \ref{modèle classique en 1D avec BC et INAR(1)},
the contagion phenomenon is embodied by the sequence $(M_k)_{k \in \N}$, which is the counting processes of the by-claims.

It is written as an INAR(1) time series, i.e for $k \geq1$, $M_k \, = \, \rho \circ M_{k-1} + N_k \,$.
\begin{itemize}
    \item $N_k$ is the source term, it stands for the first losses linked with health and environmental damages for an insured client. These first losses are assumed to occur at period $k$, at the same period where the financial compensations related to the client's civil liability are settled. 
    \item $\rho \circ M_{k-1}$ is the real contagion term, it represents the consequences of the pollution that were already causing financial losses 
    at period $k-1$, and are not completely solved or cured at period $k$. As it will be precised further, $\rho \circ M_{k-1} | M_{k-1} \ \sim \ \cB(M_{k-1}, \rho)$, so the higher the reproduction coefficient $\rho$ is, the larger the contagion is. 
\end{itemize}

In general, when $\rho$ is greater than 0.5, the risks coming from a given client have more chances to remain persistent and engender expenses in the next period than to be removed. And the closer $\rho$ is to 1, the more it must be difficult to predict the end of such a contagion event.
Precisely, for the PFAS issue, $\rho$ should be quite large, since the processes of decontamination and recovery are expected to be long because. Even with a conservative estimation relying only on intuition and not on data, we should observe $\rho \geq 0.5$.

Another fact to justify this writing and this estimation is: if one hot spot of PFAS is detected at a given place, or if some people from the same neighbourhood are affected by a disease they suspect to be due to perfluorinated compounds, then people living near have suddenly more chances to be affected and to complain about that.
\medskip

What is more, regarding the common contagion processes in ruin theory, this one seems appropriate to describe the problem. Indeed, the source of the emissions on the one hand, and their consequences on the other, are well highlighted. The sources, which refer to the clients who will benefit from financial compensations and are responsible for such pollution, are counted by the variables $(N_k)_{k \in \N}$, which are involved in the aggregate amounts of both the main claims and the by-claims. Then, the clusters of claims aroused by the health care and the curing of the environment are represented by $(M_k)_{k \in \N}$, and specifically the reproduction of the claims along the periods by the term $\rho \circ M_{k-1}$.
Alternatively, a Hawkes process would not describe faithfully enough the progressive propagation of the molecules, or the fact that the processes of health care and environmental clean-up are relatively slow. 

Even if, according to Proposition \ref{gamma_n} the sequence $(\gamma_k)_{k \in \N}$ is monotonic, the model fits if we consider that at the time origin $n=0$, strong regulatory measures are taken, and will carefully be applied. In this way, we expect to have a frequency $\gamma_0$ which is surely larger than $\frac{\lambda}{1-\rho}$. Therefore, from $n=0$, the risks engendered by the perfluorinated decrease slowly thanks to the measures taken, and aim at reaching a stationary behaviour (but it should happen after a quite long time accordingly to the context).

Thus, this model could be applied to high levels of contamination, and to a cleaning process which takes time to be handled.

\begin{rem}[Differences with a Hawkes process]
To write it once for all, we summarise here the main differences between the INAR(1) process used in this paper and the notion of Hawkes process (which was mentioned in the introduction as a type of process that may embody a contagion phenomenon).

The classical Hawkes process does have a long memory, its trajectories are usually non-monotonic, and present jumps \cite{zhu2013nonlinear}.

As we will see later, the INAR(1) process cannot be characterised as a long-memory process, because the decreasing of its "asymptotic autocorrelation" is exponential and not in the form of a negative power \cite{kouamo2011analyse}. Moreover, it is monotonic, and does not present jumps. 

Besides that, as we mentioned just above, a Hawkes process might be seen as a process with too much excitation to represent accurately the supposed slowness of the solving of the PFAS issue.
\end{rem}

\section{The Novel risk model with INAR(1) structure}
\label{section de présentation du modèle initial}

In this paper, all the random variables and stochastic processes are defined on a probability space $(\Omega, \mathcal{F}, \p)$. Moreover, $\N = \{0, 1, 2, \dots \}$ is the set of the nonnegative integers, and \\ $\N^* = \{1,2, \dots \}$ are the positive integers.
\medskip

\subsection{Presentation of the model}

We consider the unidimensional discrete-time risk model as follows:

\begin{equation}
\label{modèle classique en 1D avec BC et INAR(1)}
    R_n \ = \ u \ + \ c \, n \ - \ \sum\limits_{k=1}^n \sum\limits_{i=1}^{N_k} X_{k,i} \ - \ \sum\limits_{k=1}^n \sum\limits_{i=1}^{M_k} Y_{k,i} \qquad \text{for} \ \, n \in \N \ ,
\end{equation}

where:
\begin{itemize}
    \item $u \geq 0$ is a real number which is the initial surplus of the insurance company.
    \item $c >0$ is the constant premium rate.
    \item The independent and identically distributed (i.i.d.) sequences of positive random variables $(X_{k,i})_{(k,i) \in \N^2}$ and $(Y_{k,i})_{(k,i) \in \N^2}$ represent respectively the amounts of the main claims and the by-claims, with finite means $\mu_X$ and $\mu_Y$. We note $F$ and $G$ their respective distributions, and to simplify, we also denote $X$ and $Y$ two random variables corresponding to these distributions.
    \item The company suffers $N_k$ main claims and $M_k$ by-claims at each period $k$, $k \geq 1$, where the sequence $(N_k)_{k \in \N^*}$ is i.i.d., with $N_1 \sim \cP(\lambda)$, $\lambda >0$.
    \item The sequences $X$, $Y$ and $N$ are mutually independent. Moreover, $X$ and $Y$ are independent of the sequence $M$. The relation of dependence between $N$ and $M$ is stated as follows.
    \item For $k \geq 1$, $M_k$ is expressed in function of $M_{k-1}$, $N_k$ and a constant parameter  $\rho \in [0,1[$ according to an INAR process of order 1: 
    \begin{equation}
      M_k \ = \ \rho \circ M_{k-1} \, + \, N_k \ ,
    \end{equation}
    where $\circ$ is the binomial thinning operator such that $\rho \circ M_{k-1} = \sum\limits_{i=1}^{M_{k-1}} b_i(\rho)$. \\
    The $(b_i(\rho))_{i \in \N}$ are i.i.d. random variables with Bernoulli distribution $\cB(\rho)$, independent of $M_{k-1}$ and hence, $\ \rho \circ M_{k-1} | M_{k-1} \ \sim \ \cB(M_{k-1}, \rho)$.
    \item We settle initially $M_0 \sim \cP(\gamma_0)$, with $\gamma_0 >0$, and note $\gamma_k = \E(M_k)$ for $k \in \N$. Moreover, for a fixed $k \geq 1$, we assume $N_k$ is independent of $M_0, \dots, M_{k-1}$. We show later that with such assumptions, actually $M_{k} \sim \cP(\gamma_{k})$ for each $k$.
\end{itemize}
\smallskip

For the $k^{th}$ period, we note $A_k = \sum\limits_{i=1}^{N_k} X_{k,i}$ and $B_k = \sum\limits_{i=1}^{M_k} Y_{k,i}$, which are the respective aggregate amount of the main claims and the by-claims.
What is more, we denote by \\ $S_n \, = \, \sum\limits_{k=1}^n(A_k + B_k) \,$ the aggregate amount of the whole losses at time $n \in \N^*$,
which yields the writing:
$\ R_n \ = \ u \, + \, c \, n \, - \, S_n \ .$
\medskip

For $u \geq 0$, we define the stopping time, $\tau(u) = \inf \{ k \in \N^* \ / \ R_k < 0 \, | \, R_0 = u \}$, \\
$\tau(u) \in \N^* \cup \{+ \infty\}$. 
Then, $\psi(u) = \p(\tau(u) < \infty)$ is the ultimate ruin probability, and for $T \in \N^*$, and $\psi_T(u) = \p(\tau(u) \leq T)$ is the finite-time ruin probability at time $T$.

\subsection{Some immediate interpretations and disclaimers}

On a practical viewpoint, this structure with an INAR(1) process implies that during the $k^{th}$ period, each main claim generates directly one by-claim precisely, and then, each by-claim which has occurred at this period engenders with probability $\rho$ a new by-claim for the $(k+1)^{th}$ period.

One could underline the fact that one main claim may engenders several types of subsequent losses in the same period. In this case, we simply consider that we gather all these losses in one claim $Y_{k,i}$, which would therefore be an aggregate amount of claims (rather than a single claim).

In contrast, as suggested in the Introduction, and as it will be seen in Section \ref{section faux INAR(2)}, one could also find more likely and consistent to assume that, for such a risk model, each main claim does not necessarily engender a by-claim. 
This assumption is made through the term $N_k$ in the expression of $M_k$, and we explore in Section \ref{section faux INAR(2)} the possibility of writing this source term $\rho_2 \circ N_k$ instead. It implies that a main claim generates a by-claim only with a probability $\rho_2 \in [0,1]$. We will see that the results found in between adapt quite easily, and this is why we prefer carrying the main of our risk analysis by using the term $N_k$, since it entails less heavy mathematical formulations.
\medskip

Moreover, as a technical remark, the by-claims have only \textit{positive amounts}: it is a simplification of what could happen in practice. Indeed, in the context of cost revising for an insured loss, some main claims may be revised upward or downward. Actually, they are revised downward if some damages appear to be less costly than initially expected, and in that case, we should observe a by-claim with a negative amount.

\section{First properties of the model}
\label{section avec les premières propriétés}

\subsection{Net profit condition}

In this subsection, we outline the net profit condition related to the model.
We also provide some basic properties linked with this condition.
\medskip

We characterise first the law and the behaviour of the sequence $(M_k)_{k \in \N^*}$.

\begin{prop}[Distribution of $(M_k)_{k \in \N}$] ~\\
\label{gamma_n}
For $k \in \N$, $M_k \sim \cP(\gamma_k)$, and
$\, \gamma_k = \rho^{k} \, \left( \gamma_0 - \frac{\lambda}{1 - \rho} \right) + \frac{\lambda}{1 - \rho} \,$. \\
Since $0 \leq \rho < 1$, the sequence $(\gamma_k)_{k \in \N}$ is monotonic, such that $\underset{k \to \infty}{\lim} \gamma_k = \frac{\lambda}{1 - \rho}$, and more precisely, $M_k \overset{\cL}{\longrightarrow} M_\infty$, where 
$M_\infty \sim \cP \left( \frac{\lambda}{1 - \rho} \right)$.
\end{prop}

\begin{proof}
First, we derive the expression of $\gamma_k$ thanks to an easy induction. \\
The monotony and the limit of the sequence $(\gamma_k)_{k \in \N}$ are then quite obvious.

Moreover, the thinning operation $\circ$ is distributive regarding the operation $+$ in $\N$, and we have the particular property that for a discrete Poisson random variable, if $N \sim \cP(\lambda)$, then $\rho \circ N \sim \cP(\rho \lambda)$. Eventually, we use the stability of the Poisson distribution by the convolution of two independent variables.

We deduce comfortably the convergence of $(M_k)_{k \in \N}$. Nevertheless, we do not have something more refined than this convergence in distribution.
\end{proof}

\begin{rem}
In particular, the sequence is stationary if and only if $\gamma_0 = \frac{\lambda}{1 - \rho}$.
\end{rem}

Actually, for a classical continuous-time risk process, the net profit condition is equivalent to $\E(R_t -u) >0$ for $t >0$, which may be expressed simply if the losses are i.i.d. 
Here, the process should satisfy $\E(R_n -u) >0$ for each $n \in \N^*$.

\begin{lem}[Net profit condition]
\label{lemme condition de profit net}
We note $\gamma_{max} = \max \left( \gamma_0, \frac{\lambda}{1 - \rho} \right)$. Then, the net profit condition is:
\begin{equation}
c \  > \  \lambda \,  \mu_X \, + \, \mu_Y \, \gamma_{max} \ .
\end{equation}
\end{lem}

\begin{proof}
For all $n \in \N$, we shall have $\E(R_n -u) >0$, which is satisfied if and only if \\
$c > \lambda \cdot  \mu_X + \mu_Y \cdot \underset{k \in \N^*}{\sup} \left(\gamma_k \right)$. Due to the properties of monotony and convergence outlined previously, we deduce the formulation above.
\end{proof}

If this condition is untrue, then the ultimate ruin is almost sure, namely $\psi(u) = 1$. So, for the remainder of this paper, we assume it is true.

In particular, this condition induces a maximal value for the parameter $\rho$. Indeed, $\rho$ has to belong to the interval $[0,\rhm[$, where:
\begin{equation}
\label{equation pour rhomax}
0  < \ \rho_{max} \, = \, 1 - \frac{\lambda \mu_Y}{c - \lambda \mu_X}  \ < \ 1 \ .
\end{equation}

This expression highlights that when $\rho$ tends to be close to 1, the company should set a great premium amount $c$, or at least have a comfortable initial surplus $u$ to face the uncertainty.
\medskip

Moreover, it allows us to define a safety loading $\chi$, and in accordance with usual ruin theory frameworks, $\chi > 0$ if and only if the net profit condition is true.
\begin{equation}
\label{équation de chargement de sécurité}
\chi \ = \ \frac{\lambda \, \mu_X  \, + \, \gamma_{max} \, \mu_Y}{c} \ - \ 1   \ 
\end{equation}

\subsection{Covariances}

In this section, we present the main covariances computations regarding the model defined by Equation \ref{modèle classique en 1D avec BC et INAR(1)}.

\begin{prop}[Variances and covariances of the counting processes] ~\\
i) Since $N_k \sim\cP(\lambda)$, for $k \geq 1$, we have $\V(N_k) = \lambda$. Similarly, for $k \in \N$, $\V(M_k) = \gamma_k$. \\
ii) For $n > k \geq 1$, $\, \cov(N_n,N_k) = 0$ (by independence). Moreover, for $\, n > k \geq 0$, \\
$\cov(M_n,M_k) = \rho^{n-k} \gamma_k$. \\
iii) For $n \geq k \geq 1$, $\cov (M_n, N_k) = \rho^{n-k} \lambda$. But if $k > n$, then $\cov(M_n, N_k) = 0$.
\end{prop}
\smallskip

Consequently, it is possible to decompose the autocovariance of $(R_n)_{n \in \N}$:

\begin{prop}[Covariances of the risk process] ~\\
If $X, Y \in \cL^2$, then we can compute the following: for $n \in \N^*$, and $1 \leq l \leq n$,
\begin{equation}
\begin{aligned}
\cov(R_l,R_n) &= \, \sum\limits_{i=1}^l (\V(A_i) + \V(B_i) \, ) \, + \, 2 \sum\limits_{1 \leq i < j \leq l} \cov(B_i,B_j) + \sum\limits_{j=l+1}^n \sum\limits_{i=1}^l \cov(B_i,B_j) \\
&+ \, 2 \sum\limits_{1 \leq i \leq j \leq l} \cov(A_i,B_j)\, + \, \sum\limits_{j=l+1}^n \,  \sum\limits_{i=1}^l \cov(A_i,B_j) \quad .
\end{aligned}
\end{equation}
where
$$
\left\{
\begin{array}{ll}
        \V(A_i) = \lambda \, \mu_{(X^2)} \, ,
     &  \\ \V(B_i) = \gamma_i  \, \mu_{(Y^2)} \, ,
     & \\ \cov(B_i,B_j) = (\mu_Y)^2  \gamma_i \,  \rho^{j-i} \quad \text{for} \ 1 \leq i < j \, ,
     & \\ \cov(A_i,B_j) = \lambda  \, \mu_X  \mu_Y  \rho^{j-i} \quad \text{for} \ 1 \leq i \leq j \ .
\end{array}
\right.
$$
\end{prop}

We chose to present these results, because in similar models in ruin theory were also provided such covariances computations (see e.g. \cite{Cossette2010DiscreteTimeRM}, or \cite{ma2015bidimensional}).

We notice that $\gamma_k$, seen as a function of $\rho$, $\lambda$, or even $\gamma_0$, is an increasing function (for \\ $k \in \N$). Therefore, as a whole, we find that when the level of risk increases, the level of uncertainty increases too.

\section{Ruin probabilities with light-tailed distributions} 
\label{section ruin prob queues fines}

In this section, we assume that both the distributions of $X$ and $Y$ are light-tailed. 
It implies in particular that there exists $s >0$ such that $M_X(s) < \infty$ and $M_Y(s) < \infty$.

We display the loment-generating function (m.g.f.) of $S_n$ in the next proposition. According to \cite{ma2015bidimensional} for instance, we express it as a multivariate function, with one variable corresponding to one type of loss (thus, things are clearer). 

Further, we refer to the large deviations theory from Glynn and Whitt (see e.g. \cite{glynn1994logarithmic}) and its applications to ruin theory (see e.g. \cite{nyrhinen1998rough} and \cite{muller2001asymptotic}) to derive the explicit Lundberg equation related to, and then the existence of the adjustment coefficient for our "non-i.i.d." context.

\begin{prop}[M.g.f. in 1 dimension]~\\
\label{tfdeLaplace/FGM en 1D}
We assume $X$ and $Y$ are light-tailed.
Let $t,s \geq 0$ be real numbers, and let us assume that we have $\rho M_Y(s) < 1$, and $M_X(t) < \infty$.
Thus, for $n \in \N^*$, the moment-generating function of $S_n$ is:
\begin{equation}
\label{équation de la tfdeLaplace/FGM en 1D}
\cM_{S_n}(t,s) \, = \, \exp \left[ \gamma_0 [B_n \left( M_Y(s) \right) - 1] + \, \lambda \left(M_X(t) \, M_Y(s) \, \frac{1 - B_n \left( M_Y(s) \right) + n  (1 - \rho)}{1 - \rho M_Y(s) } - n \right) \right] \ .
\end{equation}

where $B_0(s) = 1, \ \text{then} \quad B_i(s) = \rho  s \, B_{i-1} (s) + (1 - \rho) \ $ for $i \geq 1$.
\end{prop}

\begin{cor}[Lundberg equation]
\label{équation de Lundberg}
If $X$ and $Y$ are light-tailed, then for $t >0$ such that \\ $M_X(t) < \infty$ and $\rho M_Y(t) < 1$, the Lundberg equation linked with the model is:
\begin{equation}
\label{équation de Lundberg en 1D}
c \, t \ = \ \lambda \left( \frac{M_X(t) M_Y(t) (1 - \rho)}{1 - \rho M_Y(t) } -1 \right) \ .
\end{equation}

\end{cor}

\begin{proof}
First, let us fix two real numbers $t,s >0$ with the constraint $\rho s <1$.

Then, let $n \in \N^*$ be a positive integer. We define the following probability-generating function (p.g.f.):
$$
P_n (t,s) \ = \ \E \left( t^{N_1 + \dots + N_n}  s^{M_1 + \dots + M_n} \right) \ .
$$

To compute it explicitly, let us define $B_0(s) = 1$, and then, for $i \in \N$, \\
$B_{i+1}(s) \, = \, \rho  s \, B_{i} (s) + \overline{\rho} \,$, where $\overline{\rho} = 1 -\rho$.
We have
\begin{equation}
P_n(t,s) \ = \ \left(\prod_{i=1}^n \E \left(t^{N_i}  s^{\left(N_i + \rho \, \circ \, N_i \, + \, \dots \, + \, \rho \overset{n-i}{\circ} N_i \right)} \right) \right) \, \times \, \, \E \left( s^{\left(M_0 \, + \, \rho \, \circ \,  M_0 \, + \, \dots \, + \, \rho \, \overset{n}{\circ} \, M_0 \right)} \right) \ ,
\end{equation}
where $\rho \overset{i}{\circ} N_k = \rho \circ (\rho \circ \dots (\rho \circ N_k) \dots )$ corresponds to $i$ successive thinning operations.
\medskip
 
Then, we derive the factor for $i=n-1$ to show how it works:
$$
\E \left(t^{N_{n-1}}  s^{\left(N_{n-1} + \rho \circ N_{n-1} \right)} \right) = \E \left( \E \left(t^{N_{n-1}} s^{\left(N_{n-1} + \rho \circ N_{n-1}  \right)} \, | \, N_{n-1} \right) \right) \ ,
$$
It yields
$$
\E \left(t^{N_{n-1}}  s^{\left(N_{n-1} + \rho \circ N_{n-1} \right)} \right) = \E \left( (ts)^{N_{n-1}}  \E \left( s^{\left(\rho \circ N_{n-1}  \right)} \, | \, N_{n-1} \right) \right) \ = \ \E \left( (ts)^{N_{n-1}} \left( 1 - \rho + \rho s \right)^{N_{n-1}} \right) \ ,
$$
$$ \text{and} \qquad 
\E \left(t^{N_{n-1}}  s^{\left(N_{n-1} + \rho \circ N_{n-1} \right)} \right) = \sum\limits_{k=0}^\infty t^{k} 
s^k [\rho s + \overline{\rho}]^k \, \p(N_{n-1} = k) \ .
$$
Hence, $\ \E \left(t^{N_{n-1}}  s^{\left(N_{n-1} + \rho \circ N_{n-1} + \dots \right)} \right)  = \exp [\lambda (t s B_{1}(s) - 1) ] \ ,$ as $\p(N_{n-1} = k) = \frac{e^{-\lambda} \lambda^k}{k !}$ for $k \in \N$.
Then, for $i = n-2$, the computation would start like this:
$$
\E \left(t^{N_{n-2}}  s^{\left(N_{n-2} + \rho \circ N_{n-2} + \rho \circ \rho \circ N_{n-2} \right)} \right) = \E \left( \E \left(t^{N_{n-2}}  s^{\left(N_{n-2} + \rho \circ N_{n-2} \dots \right)} \, | \, N_{n-2} , \rho \circ N_{n-2} \right) \right) \dots
$$

Hence, we obtain, for $1 \leq i \leq n$,
$\E \left(t^{N_i}  s^{\left(N_i + \rho \, \circ \, N_i + \dots\right)} \right) = \exp[\lambda (t s B_{n-i}(s) - 1)] \, $, but also 
$\E \left( s^{\left(M_0 \, + \, \rho \circ M_0 \, + \, \dots \,  \right)} \right) = \exp \left[ \, \gamma_0 (B_{n}(s) - 1) \, \right] \ .$

Therefore, 
\begin{equation}
P_n(t,s) = \exp \left( \gamma_0(B_n(s) - 1) + \lambda \left(t s \, \sum\limits_{i=0}^{n-1} B_i(s) - n \right) \right) \ .
\end{equation}
We can prove by induction that for $n \geq 1$,
$$
\sum\limits_{i=0}^{n-1} B_i(s) = \frac{1 - B_{n}(s) + n(1 - \rho)}{1 - \rho s} \ .
$$

Finally, as a definition, $\cM_{S_{n}} (t,s) = P_n \left(M_X(t), M_Y(s) \right)$, and by substitution, it yields the expression outlined above.
\bigskip

Now, from this, we derive the Lundberg equation related to the model. As mentioned, for the next part, we use some large deviation results for ruin theory.

We take here $t>0$ such that $M_X(t) < \infty$ and $\rho M_Y(t) < 1$.

For $n \in \N^*$, we define $c_{n}(t) \ = \ \ln \left(\cM_{S_{n}} (t,t) \, \right) \ - \ n  c  t$. \\
Then, the Lundberg equation corresponds to $f(t) = 0$, where $f(t) \, = \, \underset{n \to \infty}{\lim} \frac{1}{n} c_{n} (t)$.

Finally, since $\rho s < 1$, and
$\ B_n(s) = \frac{\overline{\rho} + \rho (\rho s)^n - (\rho s)^{n+1}}{1 - \rho s}$, we get $
\underset{n \to \infty}{\lim} \frac{B_n(s)}{n}  = 0 \,$,
and it yields the Lundberg equation outlined above.
\end{proof}
\smallskip

Actually, the same kind of calculus was made in \cite{Cossette2010DiscreteTimeRM}, \cite{Shi2014AnAM}, or \cite{ma2015bidimensional} for instance.

As for these papers, the asymptotic Lundberg-type result obtained regarding the so-called adjustment coefficient is only logarithmic here, i.e. $\underset{u \to \infty}{\lim} \frac{\ln(\psi(u))}{u} = - \, \kd$. 

\begin{prop}[General case with two light-tailed distributions] ~\\
\label{prop existence coeff d'ajustement modèle 1D classique}
Let us make the same assumptions as for Proposition \ref{tfdeLaplace/FGM en 1D}. \\
i) Then, when $u \to \infty$, the ultimate ruin probability is $\psi(u) \, \overset{\log}{\sim} \,  e^{-\kappa_D \, u} \, ,$ where $\kappa_D >0$ is the adjustment coefficient related with the model. \\
ii) Moreover, seen as a function of $\rho$, $\kd$ is non-increasing.
\end{prop}

\begin{proof}

i) We still refer to the large deviation theory.

Let us take $I = [0,x_0[, \, x_0 >0$, the maximal interval where the Lundberg equation is defined, that is to say $t \in I$ is equivalent to $M_X(t) < \infty$, and $\rho M_Y(t) < 1$.

We want to solve, for $t \in I$, Equation \ref{équation de Lundberg en 1D}, which can be rewritten
\begin{equation}
\label{equation de Lundberg modifiée}
f(t) \ = \ \lambda \, \frac{M_X(t) M_Y(t) (1 - \rho)}{1 - \rho M_Y(t)} - (\lambda + c t ) \ = \ 0 \ ,
\end{equation}
and if at least one solution exists, the so-called adjustment coefficient would be the smaller solution.

$f$ is a convex function. Moreover, $f(0) =0$, and $f'(0) = \lambda \left(\mu_X + \frac{\mu_Y}{1-\rho} \right) - c$, so $f'(0) < \, 0$ thanks to the net profit condition.

Moreover, by the definition of $I$ and $x_0$, $\lim\limits_{t \to x_0^-} \frac{M_X(t) }{1 - M_Y(t) \, \rho}= + \infty$. \\
Therefore, $\lim\limits_{t \to x_0^-} f(t) = + \infty$.
We deduce that $f(t) < 0$ in a neighbourhood of 0, and regarding the limit of $f$ in $x_0$, thanks to the intermediate value theorem, Equation \ref{equation de Lundberg modifiée} has at least one solution, and $\kd$ exists.
\bigskip

ii) Let us see $f$ as a function of $\rho$. $f$ is (at least) $\cC^1$ according to this parameter, and
\begin{equation}
\frac{\partial f}{\partial \rho} (t,\rho) \ = \ \lambda \frac{M_X(t) M_Y(t) \left[M_Y(t) - 1 \right]}{(1- \rho M_Y(t) )^2} \ .
\end{equation}
So $\frac{\partial f}{\partial \rho} (t,\rho) \geq 0$, since for each $t \geq 0$, $M_Y(t) \geq 1$.

According to the previous reasoning, it means that if $\rho_1 \leq \rho_2$, then $f(t, \rho_1) \leq f(t,\rho_2)$ for every $t \geq 0$, and therefore $\kd(\rho_1) \geq \kd(\rho_2)$. In other words, $\kd$ is a nonincreasing function of $\rho$.
\end{proof}

\begin{rem}
Actually, we have to be precise and distinguish between the possible types of light-tailed distributions. This previous property is true if the distributions of $X$ and $Y$ are such that $M_X(t) < \infty$ for $t \in [0, M[$, where $M \in ]0,\infty]$ (and the same thing for $Y$), but it would not be necessary true if we had $M_X(t) < \infty$ for $t \in [0, M]$ with $0 < M < \infty$ (or similarly for $Y$). However, these distributions are rare, so for the rest of the paper and the following other adjustment coefficient analyses, we do not take into account this kind of distribution. 
\end{rem}

Then, it is possible to derive the explicit expression of $\kd$ when the amounts of the two kinds of losses both follow an exponential distribution.

\begin{prop}[With 2 exponential distributions] ~\\
\label{ruine1Dcasexpdouble}
Let us assume that $X \sim \operatorname{Exp}(\eta)$, and $Y \sim \operatorname{Exp}(\delta)$, with $\eta, \delta >0$. \\
Then, when $u \to \infty$, we have the asymptotic ruin probability:
\begin{equation}
\label{kappa d'ajustement modèle original (dépendant) 1D}
\psi(u) \, \overset{\log}{\sim}  \, e^{-\kappa_D \, u}  \ , \quad
\kappa_D \ = \ \frac{\eta + \delta(1 - \rho)}{2} \ - \ \frac{\lambda + \sqrt{c^2 [\eta - \delta (1-\rho)]^2 + \lambda^2 + 2 \lambda c [\eta + \delta (1 - \rho)]}}{2c} \ . 
\end{equation}
\end{prop}

\begin{proof}
Since $M_X(t) = \frac{\eta}{\eta - t}$, and
$M_Y(t) = \frac{\delta}{\delta - t}$, the Lundberg equation is here equivalent to 
\begin{equation}
\label{equation de Lundberg en 1D avec 2 lois exp}
a_2 t^2 + a_1 t + a_0 \ = \ 0 \quad \text{where} \quad 
\left\{
\begin{array}{ll}
        a_0 \, = \, c \eta \delta (1- \rho) \,  - \,  \lambda (\eta + \delta (1 - \rho) ) \ ,
     &  \\ a_1 \, = \,  \lambda - c [\eta + \delta (1 - \rho)] \ ,
     & \\ a_2 \, = \, c \ .
\end{array}
\right.
\end{equation}
We have $a_1 < 0 , \ a_0 >0 \, $ thanks to the net profit condition, and of course $a_2 >0$. \\
We also define:
$$ \Delta_D \ = \ a_1^2 - 4 a_0 a_2 \ = \ c^2 [\eta - \delta (1-\rho)]^2 + \lambda^2 + 2 \lambda c [\eta + \delta (1 - \rho)] \ .$$
Finally, $\kappa_D$ is the smallest positive root of the second degree equation $a_2 t^2 + a_1 t + a_0 = 0$, i.e.
$\kappa_D \ = \ \frac{-a_1 - \sqrt{\Delta_D}}{2 \, a_2} \ ,$
which yields the expression outlined.
\end{proof}

The expression of $\kd$ is quite heavy, so we would like to perceive it better, and in particular have bounds of the related incurred risk by an insurance company.

We can check that there is no redundancy in the following results. For example, the first upper bound is more refined when $\eta$ and $\delta(1 - \rho)$ are close, whereas the second one is thinner when it is not the case. 

A lower bound of $\kd$ would correspond to an upper bound of the risk incurred, but actually, such bounds were more difficult to derive.

\begin{prop}[Bounds of $\kappa_D$]
\label{majoration kappa 1D}
We have the upper bounds:
\begin{equation}
\label{équation avec majorants de kappa_D}
\kappa_D \ < \ \frac{\eta + \delta (1-\rho)}{2} \, - \, \frac{\lambda}{c} \qquad ; \qquad
\kappa_D \ < \ \min (\eta , \delta (1-\rho) \, ) \, - \, \frac{\lambda}{2c} \ .
\end{equation}

\end{prop}

\begin{proof}
These bounds can be deduced by "direct" computations. For the first bound, we can observe that $\sqrt{\Delta_D} > \lambda \, $, and for the second, $\, \sqrt{\Delta_D} > c|\eta - \delta(1-\rho)| \, .$ 

With similar considerations, it is possible to draw some other comparisons: for instance, by noticing that $\sqrt{\Delta_D} > |\lambda + c(\eta - \delta(1-\rho))|\,$ (we provide more details just after this proof).
\end{proof}

With only main claims, no by-claims, and an INAR(1) process counting the main claims, \cite{Cossette2010DiscreteTimeRM} found the following adjustment coefficient (with an adaptation of the notations):
\begin{equation}
\kappa_1 \, = \, \delta(1-\rho) - \frac{\lambda}{c} \ .  
\end{equation}
Since $\sqrt{\Delta_D} > |\lambda + c(\eta - \delta(1-\rho))|\,$, we get $\kd < \kappa_1$, hence our model from Equation \ref{modèle classique en 1D avec BC et INAR(1)} is more risky (which was quite intuitive).

Moreover, if $X=0$ a.s., we would find $\kd = \kappa_1$. And if, in addition, we compare the respective expressions, it seems that $\kd$ is a fair generalisation of $\kappa_1$ with another type of loss.

\section{Models with softened dependence}
\label{section de comparaison des modèles indépendants}

From our initial risk model in Equation \ref{modèle classique en 1D avec BC et INAR(1)}, we define two other similar models, here we soften the dependence assumptions on the counting processes: thus, we present a \textit{"half-independent"} model, and a \textit{"purely independent"} model. Moreover, here we call the original model in Equation \ref{modèle classique en 1D avec BC et INAR(1)} the \textit{"dependent model"}, and we rewrite it as $R^D$ instead of $R$. 

We express some of their risk properties, in particular their adjustment coefficients, and then we compare these three models.

\subsection{Half-independent model}

The half-independent model is :
\begin{equation}
\label{modèle semi-indépendant en 1D}
R^S_n \ = \ u \ + \ c \, n \ - \ \sum\limits_{k=1}^n \sum\limits_{i=1}^{N_k} X_{k,i} \ - \ \sum\limits_{k=1}^n \sum\limits_{i=1}^{M^S_k} Y_{k,i} \qquad \text{for} \ n \in \N \ .
\end{equation}
We keep most of the assumptions of the dependent model, but what changes is that $M^S_k$ is no longer written as an INAR(1) model, but instead $M^S_k = N_k + O_k$ for $k \geq 1$.

Thus, we maintain the source term $N_k$, but we replace the contagion term by a variable $O_k$, defined this way: $O_k \sim \cP(\gamma_k - \lambda)$ for $k \in \N^*$, and $(O_k)_{k \in \N^*}$ is a sequence of independent variables, which is independent of the sequences $N$, $X$ and $Y$.

In particular, it yields that $M^S_k \sim \cP(\gamma_k)$ for each $k \in \N^*$. Hence, $M^S_k$ has the same distribution as $M^D_k$.
Actually, the sequences $(N_k)_{k \in \N^*}$ and $(M^S_k)_{k \in \N^*}$ are still dependent, but now, the variables $M^S_k$ are mutually independent. Specifically, for $k \in \N^*$, $M^S_k$ and $N_k$ are dependent, but $M^S_k \perp N_j$ as soon as $j \neq k$.

\begin{prop}[Adjustment coefficient of the half-independent model] ~\\
\label{model1D semi-indépendant cas exp double}
We assume that the losses have the same assumptions as in Proposition \ref{ruine1Dcasexpdouble}, that is to say \\ $X \sim \operatorname{Exp}(\eta)$, and $Y \sim \operatorname{Exp}(\delta)$, with $\eta, \delta >0$. \\
Therefore, the half-independent model satisfies $\psi_S(u) \, \overset{\log}{\sim} \, e^{- \ks \, u}$, with the following adjustment coefficient:
\begin{equation}
\label{kappa d'ajustement modèle semi-indépendant 1D}
\ks \ = \ \frac{\eta + \delta }{2} \, - \, \frac{\lambda + \sqrt{4c \lambda \delta (1-\rho)^2 + [\lambda + c(1-\rho)(\eta - \delta)]^2}}{2c(1-\rho)} \ .
\end{equation}
\end{prop}

\begin{proof}
By writing $\, S^S_n \ = \ \sum\limits_{k=1}^n \sum\limits_{i=1}^{N_k} X_{k,i} \, + \, \sum\limits_{k=1}^n \sum\limits_{i=1}^{M^S_k} Y_{k,i} \ $ the cumulative amount of losses at time $n \in \N^*$, we find its m.g.f. similarly as for Proposition \ref{ruine1Dcasexpdouble}:
\begin{equation}
    \cM_{S^S_n}(t,s) = \exp \left(n \lambda M_Y(s)(M_X(t)-1) \ + \ (M_Y(s)-1) \sum\limits_{i=1}^n \gamma_i \right) \ .
\end{equation}
As before, according to large deviations results, we deduce the Lundberg equation:
\begin{equation}
ct \, = \, \lambda M_Y(t)(M_X(t) - 1) \, + \, \frac{\lambda}{1 - \rho} (M_Y(t) -1)\ .
\end{equation}
We have here $M_X(t) = \frac{\eta}{\eta -t}$, and $M_Y(t) = \frac{\delta}{\delta - t}$, which leads to the second degree equation
$a_2  t^2 + a_1  t + a_0 = 0$, where:
\begin{equation}
\left\{
\begin{array}{ll}
        a_0 = c \eta \delta (1- \rho) - \lambda (\eta + \delta (1 - \rho) \, ) \ ,
     &  \\ a_1 = \lambda - c(1 - \rho) (\eta + \delta) \ ,
     & \\ a_2 = c (1 - \rho) \ .
\end{array}
\right.
\end{equation}
$a_2 >0$ is obvious, and the fact that $a_1 <0$ and $a_0>0$ is a consequence of the net profit condition. 
Then, we deduce the adjustment coefficient $\ks$ as the smallest positive solution of this equation.
\end{proof}

\subsection{Purely independent model}

The "purely independent" model is:
\begin{equation}
\label{modèle purement indépendant en 1D}
R^I_n \ = \ u \ + \ c \, n \ - \ \sum\limits_{k=1}^n \sum\limits_{i=1}^{N_k} X_{k,i} \ - \ \sum\limits_{k=1}^n \sum\limits_{i=1}^{M^I_k} Y_{k,i} \qquad \text{for} \ n \in \N \ .
\end{equation}
We keep most of the assumptions of the dependent model, but this time, we simply set a sequence of independent variables $(M^I_k)_{k \in \N^*}$ such that $M^I_k \sim \cP(\gamma_k)$ for each $k \in \N^*$.
Moreover, we assume that the counting processes $(N_k)_{k \in \N^*}$ and $(M^I_k)_{k \in \N^*}$ are independent, and that $M^I$ is also independent of $X$ and $Y$. In short, here all the sequences of variables $X$, $Y$, $N$ and $M^I$ are mutually independent.

\begin{prop}[Adjustment coefficient of the purely independent model]~\\
\label{model1D purement indep cas exp double}
We still assume that $X \sim \operatorname{Exp}(\eta)$, and $Y \sim \operatorname{Exp}(\delta)$, with $\eta, \delta >0$. \\
Therefore, this model satisfies $\psi_I(u) \, \overset{\log}{\sim} \, e^{- \ki \, u}$, with:
\begin{equation}
\label{kappa d'ajustement modèle rendu indépendant 1D}
\ki \ = \ \frac{\eta + \delta }{2} \, - \, \frac{\lambda(2 - \rho)+\sqrt{4 \lambda^2(1-\rho) + [\lambda \rho + c(1-\rho)(\eta - \delta)]^2}}{2c(1-\rho)} \ .
\end{equation}
\end{prop}

\begin{proof}
It is the same kind of proof as for Proposition \ref{kappa d'ajustement modèle semi-indépendant 1D}, but here we have:
\begin{equation}
    \cM_{S^I_n}(t,s) = \exp \left(n\lambda (M_X(t)-1) + (M_Y(s)-1) \sum\limits_{i=1}^n \gamma_i \right) \ .
\end{equation}
Then, the Lundberg equation is
\begin{equation}
ct \, = \, \lambda(M_X(t) - 1) \, + \, \frac{\lambda}{1 - \rho} (M_Y(t) -1) \ .
\end{equation}
It leads to
$a_2  t^2 + a_1  t + a_0 = 0$, where:
\begin{equation}
\left\{
\begin{array}{ll}
        a_0 = c \eta \delta (1- \rho) - \lambda (\eta + \delta (1 - \rho) \, ) \ ,
     &  \\ a_1 = \lambda(2 - \rho) - c(1 - \rho) (\eta + \delta) \ ,
     & \\ a_2 = c (1 - \rho) \ .
\end{array}
\right.
\end{equation}
$a_2 >0$ is obvious, then $a_1 <0$ and $a_0>0$ thanks to the net profit condition. 
Then, we deduce $\ki$ as the smallest positive solution of this equation.
\end{proof}

\subsection{Some first comparisons of the three models}

Here we derive the main result of this section, namely, we understate the risk incurred if we underestimate the level of dependence of the contagion process (at least for this particular case of exponential distributions).
In relation with the PFAS issue, it should entice companies who are covering such risks to assess carefully the potential underlying dependence.
\medskip

Before that, we propose to compare the means and variances of the aggregate amount of losses for the 3 models, which we note here $S_n^D$, $S_n^S$ and $S_n^I$.

\begin{prop}[Comparison of the means and variances] ~\\
We have 
$\E(S_n^I) = \E(S_n^S) = \E(S_n^D) = n \lambda \mu_X + \mu_Y  \sum\limits_{i=1}^n \gamma_i$. \\
However, if $X, \, Y \in \cL^2$, $ \V(S_n^I) \, < \, \V(S_n^S) \, < \, \V(S_n^D) \,$, and \\
$\underset{n \to \infty}{\lim} [\V(S_n^D) - \V(S_n^S)]  \, = \,  \underset{n \to \infty}{\lim} [(\V(S_n^S) - \V(S_n^I)] \, = \, + \infty \, . $ \\
Indeed, $ \V(S_n^S) \, - \, \V(S_n^I) \ =  \ 2n\lambda \mu_X \mu_Y  \ , \ $ and \\
$\V(S_n^D) - \V(S_n^S) \ = \ 2 \sum\limits_{j=2}^n \sum\limits_{i=1}^{j-1} (\mu_Y)^2 \gamma_i \rho^{j-i} \ + \ 2 \mu_X \mu_Y \, \frac{\rho \lambda}{(1-\rho)^2} \left[ n(1-\rho) - (1-\rho^n) \right] \ .$
\end{prop}

\begin{proof} 
Computing the variances yields:
$ \V(S_n^I) \ = \ n \lambda \mu_{(X^2)} \ + \ \mu_{(Y^2)} \sum\limits_{i=1}^n \gamma_i \ , \ $ and \\
$ \V(S_n^S) \ =  \  n \lambda \mu_{(X^2)} \ + \ \mu_{(Y^2)} \sum\limits_{i=1}^n \gamma_i \ + \ 2n\lambda \mu_X \mu_Y  \ , \ $ and \\
\small{$\V(S_n^D) \ = \ n \lambda \mu_{(X^2)} \ + \ \mu_{(Y^2)} \sum\limits_{i=1}^n \gamma_i \ + \ 2 \sum\limits_{j=2}^n \sum\limits_{i=1}^{j-1} (\mu_Y)^2 \gamma_i \rho^{j-i} \ + \ 2 \mu_X \mu_Y \, \frac{\lambda}{(1-\rho)^2} \left[ n(1-\rho) - \rho(1-\rho^n) \right] \ .$}
\end{proof}
\medskip

After that, we are interested in comparing the adjustment coefficients related to this double exponential case. We first outline a quite intuitive result:

\begin{prop}[Variations of $\ks$ and $\ki$ according to $\rho$] ~\\
Seen as functions or $\rho$, $\ks$ and $\ki$ are non-increasing.
\end{prop}

\begin{proof}
We could prove it the same way as point ii) from Proposition \ref{prop existence coeff d'ajustement modèle 1D classique} for a general case regarding the models $(R^S_n)_{n \in \N}$ and $(R^I_n)_{n \in \N}$. Otherwise, since for this particular exponential case we know the exact expressions, we may compute directly and use the derivatives $\ks'(\rho)$ and $\ki'(\rho)$.
\end{proof}

Then, we rank the adjustment coefficients:

\begin{prop}[Comparison of the 3 adjustment coefficients] ~\\
\label{comparaison des 3 coefficients d'ajustement}
We keep the assumptions of Proposition \ref{ruine1Dcasexpdouble}: $X \sim \operatorname{Exp}(\eta)$, and $Y \sim \operatorname{Exp}(\delta)$, $\eta, \delta >0$.
Therefore, $\, \kd(0) = \ks(0) < \ki(0)$, and then, for $\rho \in ]0,\rho_{max}[$,
\begin{equation}
\kappa_D(\rho) \ < \ \kappa_S(\rho) \ < \ \kappa_I(\rho) \ .
\end{equation}
However, when $\rho \overset{<}{\to} \rho_{max}$, we have $\underset{\rho \to \rho_{max}}{\lim} \kappa_D(\rho) \ = \ \underset{\rho \to \rho_{max}}{\lim} \kappa_S(\rho) \ = \ \underset{\rho \to \rho_{max}}{\lim} \kappa_I(\rho) = 0$.
\end{prop}

\begin{proof}
First of all, regarding the expressions, we see $\ki$, $\ks$ and $\kd$ as continuous functions according to the positive parameters $c, \rho, \eta, \delta$, and $\lambda$.

By extending the definition domains to $\rho = \rhm$ (even if the net profit condition is not fulfilled for this value), we get $\ki(\rhm) = \ks(\rhm) = \kd(\rhm) = 0$.
It yields the limits outlined, by continuity.
\medskip

Let us compare $\kd$ and $\ks$. First, we find easily $\ks(0) = \kd(0)$, and after that, we consider $\ks(\rho) = \kd(\rho)$ for $\rho \in ]0,\rhm]$.

We square twice the equations to get rid of the square roots involved in the expressions of $\kd$ and $\ks$.
By rearranging the terms, we have a fourth-degree polynomial equation with $\lambda$ and $c$ as variables.
After some heavy computations and simplifications, we find that $\ks(\rho) = \kd(\rho)$ implies $\rho^2 \, \lambda(\eta + \delta(1-\rho)) = \rho^2 \, c\eta \delta(1-\rho)$.
If we assume $\rho \neq 0$, it induces \\
$\lambda(\eta + \delta(1-\rho)) = c\eta \delta(1-\rho)$, which entails $\rho = \rhm$.
Therefore, $\ks(\rho) = \kd(\rho)$ has no solution in $]0,\rhm[$.

However, we need an additional argument to justify that $\ks(\rho) > \kd(\rho)$ for $\rho \in ]0, \rhm[$. We compare the values of the derivatives $\ks'(\rho)$ and $\kd'(\rho)$ in 0, and we find that $\ks'(0) = \kd'(0)$ implies $\eta (c \delta - \lambda)^2 + \lambda c \delta (\delta + \eta) = 0$.
It is never true, as $\lambda c \delta (\delta + \eta) >0$, and $\eta (c \delta - \lambda)^2 \geq 0$. Moreover, $\ks'(0)$ and $\kd'(0)$ are continuous, and for $\lambda = \eta = \delta = 1$, and $c = 5$, we obtain $\ks'(0) > \kd'(0)$, thus (for the same set of values), there exists $0 < \e < \rhm$ such that \\ $\ks(\rho) > \kd(\rho)$ for $\rho \in ]0,\e[$. By continuity, $\ks(\rho) > \kd(\rho)$ for $\rho \in ]0,\rhm[$, and for every $(c,\eta, \delta, \lambda)$ (such that the net profit condition is satisfied).
\medskip

Let us now compare $\ki$ and $\ks$. 

First of all, we find $\ki(0) > \ks(0)$. For that, we solve the equation $\ki(0) = \ks(0)$, and we apply the same kind of comparison than for $\ks'(0)$ and $\kd'(0)$.

Then, with similar calculus, the equation $\ks(\rho) = \ki(\rho)$ for $\rho \in ]0,\rhm[$ leads to \\ $\lambda(\eta + \delta(1-\rho)) = c\eta \delta(1-\rho)$.
Once again, the only solution on $]0,\rhm]$ is $\rho = \rhm$.
Finally, $\ki(\rho) > \ks(\rho)$ for every $\rho \in [0,\rhm[$, and every $(c,\eta, \delta, \lambda)$.
\end{proof}

An example of work in ruin theory where such a comparison between the Lundberg coefficients is made when we lighten some dependence assumptions is the one from \cite{wang2005correlated}, which describes a multivariate model where a claim occurring in a line of business can affect the other lines.
\bigskip

In addition, we can derive lower bounds regarding $\ki$. It is possible since the variables $M_k^I$ are mutually independent for $k \geq 1$, and independent of the sequence $(N_k)_{k \in \N}$, but not for $\kd$ because the $M^D_k$ are dependent (for $k \geq 0$). Similarly, it is neither possible for $\ks$, since $(M^S_k)_{k \in \N}$ and $(N_k)_{k \in \N}$ are dependent. 

The corresponding properties used to prove the following proposition but also some further results, and according to which it is not possible to extend that to the models $(R^D_n)_{n \in \N}$ and $(R^S_n)_{n \in \N}$, are Theorems 1.A.3 and 1.A.4 from \cite{shaked2007stochastic}. 

\begin{rem}
We also acknowledge that if the stochastic comparisons were possible, we would expect to have $R^D_n \leq_{st} R^S_n \leq_{st} R^I_n$ for $n \in \N$, or at least to derive in general \\
$\psi_I(u) \leq \psi_S(u) \leq \psi_D(u)$. However, we were only able to establish such a risk comparison for the adjustment coefficients.
\end{rem}

\begin{prop}[Lower bounds of $\ki$]
\label{bornes inf de kappa_I cas exp double}
We have these lower bounds for $\ki$, which actually correspond to an upper bound of the incurred risk:
\begin{equation}
\kappa_I \ \geq  \ \eta \wedge \delta \,  - \, \frac{\lambda (2 - \rho)}{c(1 - \rho)} \ \qquad ; \qquad \kappa_I  \ \geq \  \frac{\eta \wedge \delta}{2} - \frac{\lambda}{c(1 - \rho)} \quad .
\end{equation}
\end{prop}

\begin{proof}
i) Let us prove the first bound. We assume first that the sequence $(\gamma_k)_{k \in \N^*}$ is decreasing.
\smallskip

Let us fix a positive integer $n_0 \geq 1$. We define the sequence $(L_k(n_0))_{k \in \N^*}$ such that \\ $L_k(n_0) \sim \cP(\gamma_k)$ for $1 \leq k \leq n_0$, and $L_k(n_0) \sim \cP(\gamma_{n_0})$ for $k \geq n_0$. Moreover, we assume that $(L_k(n_0))_{k \in \N^*}$ is independent of the sequences $(N_k)_{k \in \N^*}$, $(M^I_k)_{k \in \N^*}$, $(X_{k,i})_{(k,i) \in \N^2}$ and $(Y_{k,i})_{(k,i) \in \N^2}$, and that the variables $L_k(n_0)$ are mutually independent for $k \in \N^*$.

Thus, $M_k \leq L_k(n_0)$ for a fixed $k \in \N^*$. So, thanks to \cite{shaked2007stochastic},
$$
\sum\limits_{i=1}^{N_k} X_{k,i} + \sum\limits_{i=1}^{M^I_k} Y_{k,i} \quad \leq_{st} \quad \sum\limits_{i=1}^{N_k} X_{k,i} + \sum\limits_{i=1}^{L_k(n_0)} Y_{k,i} \quad \leq_{st} \quad \sum\limits_{i=1}^{N_k} Z_{1,k,i} + \sum\limits_{i=1}^{L_k(n_0)} Z_{2,k,i} \ .
$$
where the sequence $(Z_{j,k,i})_{(k,i) \in \N^2, 1 \leq j \leq 2}$ is i.i.d., following a distribution $\operatorname{Exp}(\eta \wedge \delta)$, and independent of $N$, $M^I$, $X$, $Y$, and $L(n_0)$. \\
This i.i.d. assumption provides $\sum\limits_{i=1}^{N_k} Z_{1,i,k} + \sum\limits_{i=1}^{L_k(n_0)} Z_{2,i,k} \ \sim \sum\limits_{i=1}^{N_k + L_k(n_0)} Z_{1,i,k} \ .$ \\
Then, by independence, $ N_k + L_k(n_0) \, \sim \, \cP(\lambda + \gamma_{n_0})$ for $k \geq n_0$.

We define the risk process: $P_n(n_0) = u + c \cdot n - \sum\limits_{k=1}^n \sum\limits_{i=1}^{N_k + L_k(n_0)} Z_{1,k,i} \ .$
By the same kind of proof as for Proposition \ref{tfdeLaplace/FGM en 1D}, we obtain this Lundberg equation:
\begin{equation}
0 = (\lambda + \gamma_{n_0})(M_Z(t) - 1) - ct \quad .
\end{equation}
As $M_Z(t) = \frac{\eta \wedge \delta}{\eta \wedge \delta - t}$, the equation has a unique solution, which yields:
$$
\psi_I(u) \ \leq \ \psi_{P(n_0)}(u) \ \overset{\log}{\sim} \ e^{-\kappa(n_0) \, u} \ , \quad \kappa(n_0) = (\eta \wedge \delta) \, - \frac{\lambda + \gamma_{n_0}}{c} \ .
$$
Thus, we infer that $\kappa(n_0) \leq \kappa_I$.

Since it is true for each $n_0 \geq 1$, when $n_0 \to \infty$, it yields $\kappa_I \geq (\eta \wedge \delta) - \frac{\lambda(2-\rho)}{c(1 - \rho)} \ .$

And, if $(\gamma_k)_{k \in \N^*}$ is nondecreasing, we may simply take $L_k \sim \cP \left( \frac{\lambda}{1 - \rho} \right)$.
\medskip

ii) We practice the same kind of reasoning for the second lower bound, but instead, we use the stochastic inequality 
$ \sum\limits_{i=1}^{N_k} X_{k,i} + \sum\limits_{i=1}^{M^I_k} Y_{k,i} \quad \leq_{st} \quad 2 \, \sum\limits_{i=1}^{L_k(n_0)} Z_{k,i} \ $ for a fixed $k \in \N^*$, 
where the $(Z_{k,i})_{(k,i) \in \N^2}$ are i.i.d., following a distribution $\operatorname{Exp}(\eta \wedge \delta)$. \\
Then, the corresponding Lundberg equation is: 
\begin{equation}
\ \gamma_{n_0} \, (\cM_Z(2t) -1) \ - \ ct \ = \ 0 \, .
\end{equation}
It provides $\kappa_I \, \geq \, \kappa(n_0) = \frac{\eta \wedge \delta}{2} - \frac{\gamma_{n_0}}{c} \ .$
Then, as $n_0 \to \infty$,
$\kappa_I \, \geq \, \frac{\eta \wedge \delta}{2} - \frac{\lambda}{c(1-\rho)} \ .$
\end{proof}

\begin{rem}
Actually, thanks to such comparisons, we might also be able to prove that, whatever are the fixed distributions for $X$ and $Y$, then for $\rho_1 \leq \rho_2$, $R^D_n(\rho_2) \leq_{st} R^D_n(\rho_1)$, or at least $\psi_D(u, \rho_1) \leq \psi_D(u, \rho_2)$. The same problem exists for $(R^S_n)_{n \in \N}$. However, it is true for $(R^I_n)_{n \in \N}$, and we write it just after as a result.
\end{rem}

\begin{prop}[Variations of $\psi_I(u)$ according to $\rho$] ~\\
For any distributions assigned to $X$ and $Y$, and for two real numbers $0 \leq \rho_1 \leq \rho_2 < \rhm$, then, for every $u \geq 0$, $\, \psi_I(u, \rho_1) \leq \psi_I(u, \rho_2)$.
\end{prop}

We could prove it quite easily thanks to suitable stochastic comparisons.

\section{Ruin probabilities with heavy-tailed distributions}
\label{section de résultats de ruine avec queue lourde}

In this section, we propose some results related to heavy-tailed distributions.
We continue to draw comparisons between the main model (Equation \ref{modèle classique en 1D avec BC et INAR(1)}) and the two models with softened dependence (Equations \ref{modèle semi-indépendant en 1D} and \ref{modèle purement indépendant en 1D}). Actually, the most precise results are obtained for the purely independent model.

Regarding the PFAS issue, we could assume the distribution of the losses would likely be heavy-tailed, especially the distribution of the by-claims (yet, this should be checked with reliable data). Indeed, the risks carried by such a pollution should be high, and the health and environmental damages should represent huge amounts: as an example, in the European Union, the annual related health costs are between 52 and 84 billion euros, and the expenses to remove these molecules from drinking water are around 238 billion euros.

\subsection{Asymptotics for subexponential and Pareto distributions}

An example of a heavy tail is the Pareto distribution. There are several ways to define this distribution, and we choose the following:

\begin{defi}[Pareto distribution]
\label{définition de la loi de Pareto}
$X \sim \operatorname{Par}(\alpha, \theta)$ with $\alpha, \theta >0$ if for $x \in \R_+$, \\
$ \ \overline{F_X}(x) = \left( \frac{\theta}{\theta + x}\right)^\alpha $. The mean is finite if and only if $\alpha >1$, and in that case, $\mu_X = \frac{\theta}{\alpha -1}$.
\end{defi}

More specifically, the Pareto distribution belongs to the class of subexponential distributions, which is itself a subclass of the heavy-tailed distribution.

\begin{prop}[Finite-time ruin probabilities with regular variations] ~\\
\label{1Ddeuxqueueslourdes}
Let us consider the original model from Equation \ref{modèle classique en 1D avec BC et INAR(1)}. \\
We assume that $F \in \cR_{-\alpha}$, and $G \in \cR_{- \beta}$, where $\alpha, \beta >1$. For $T \in \N^*$, \\
i) If $\alpha < \beta$, then 
$\psi_T(u) \in \cR_{-\alpha}$ (seen as a function of $u$). \\
ii) If $\alpha > \beta$, therefore $\psi_T(u) \in \cR_{-\beta}$. \\
iii) If $\alpha = \beta$, we get naturally $\ \psi_T(u) \in \cR_{-\alpha}$.
\end{prop}

\begin{proof}
We first look at the particular case where $X$ and $Y$ are Pareto distributions: $X \sim \operatorname{Par}(\alpha, \theta)$ and $Y \sim \operatorname{Par}(\beta, \eta)$ (with $\alpha, \beta >1$ and $\theta, \eta >0$).

We apply Lemma X.2.2 of \cite{asmussen2010ruin}, which yields, for  each $k \in \N^*$, $\overline{F_{A_k}}(x) \sim \lambda \overline{F}(x)$ and \\
$\overline{F_{B_k}}(x) \sim \gamma_k \overline{G}(x)$.

On the other hand, we can write $\psi_T(u) = \p \left( \underset{1 \leq j \leq T}{\max} \left( \sum\limits_{k=1}^j (A_k + B_k) - cj \right) > u \right)$. Thus, we have
$\ \p \left( \underset{1 \leq j \leq T}{\max} \sum\limits_{k=1}^j (A_k + B_k) > u + c T \right) \, \leq \, \psi_T(u) \, \leq \, \p \left( \underset{1 \leq j \leq T}{\max} \sum\limits_{k=1}^j (A_k + B_k) > u \right) \ . $

Since $T$ is fixed, as $u \to \infty$, $\psi_T(u) \sim \p \left( \underset{1 \leq j \leq T}{\max} \sum\limits_{k=1}^j (A_k + B_k) > u \right) $.
Moreover, by positivity, $\underset{1 \leq j \leq T}{\max} \sum\limits_{k=1}^j (A_k + B_k) = \sum\limits_{k=1}^T (A_k + B_k)$.
Then, we are able to deduce the explicit equivalents with these Pareto distributions, by using in particular Proposition X.1.9 and Corollary X.1.11 from \cite{asmussen2010ruin}.

For the case i), since $\overline{G}(x) = o(\overline{F}(x))$, it gives $\psi_T(u) \sim (\theta^\alpha  T \lambda ) u^{- \alpha}$.

For ii), if $Y \sim \operatorname{Par}(\beta, \eta)$, it gives $\psi_T(u) \ \sim \ \left( C_T \, \eta^\beta \right) u^{-\beta}  \,$, where
$$C_T \ = \  \sum\limits_{k=1}^T \gamma_k \ = \  \left( \frac{\lambda}{(1-\rho)^2} [T(1-\rho) - \rho(1-\rho^T)]\, + \, \frac{\rho \gamma_0 (1-\rho^T)}{1 - \rho}
\right) \ .$$

Eventually, for iii), $\, \psi_T(u) \sim \left(T \lambda \theta^\alpha \, + \, C_T \, \eta^\alpha \right)\, x^{- \alpha} \ .$

These three points i), ii) and iii) may be generalised to the classes $\cR_{-\alpha}$ and $\cR_{-\beta}$ by adapting the proof to appropriate slowly varying functions $L_1$ and $L_2$, such that \\
$\overline{F}(x) = L_1(x) x^{-\alpha}$, and $\overline{G}(x) = L_2(x) x^{-\beta}$ for $x \geq 0$.
\end{proof}

This result could be adapted to the two models with softened dependence.
\bigskip

Now, let us introduce a classical continuous-time Cramér-Lundberg risk process: \\
$R_c(t) = u +ct - \sum\limits_{i=1}^{M(t)} Y_i$, where $u \geq 0$, $c >0$, the $(Y_i)_{i \in \N^*}$ are positive i.i.d. random variables with mean $\mu >0$, and $M \sim \cP o(\lambda)$ is a Poisson process.
We denote by $\psi_c(u)$ its ultimate ruin probability.

In particular, if the stationary excess distribution of $Y_1$ is subexponential, then, the asymptotic ruin probability when $u \to \infty$ is
\begin{equation}
\label{equivalent continu loi sous-exp}
\psi_c(u) \, \sim \, \frac{\lambda}{c - \lambda \mu} \int_u^\infty \overline{F_Y}(x) \, dx
\end{equation}
In particular, if $Y_i = V_i + W_i$ is the sum of two independent variables (for instance, a main claim and a by-claim), we write 
\begin{equation}
\label{equivalent continu loi sous-exp avec by-claims}
\psi_c(u) \, \sim \, \frac{\lambda}{c - \lambda (\mu_V + \mu_W)} \int_u^\infty \left( \overline{F_V}(x) + \overline{F_W}(x) \right) \, dx \ .
\end{equation}

We can define the transcription of $R_c$ in discrete time: for $n \in \N$, \\
$Q_n \ = \ u \, + \, cn \, - \, \sum\limits_{k=1}^n \sum\limits_{i=1}^{N_k} X_{k,i} \, , \ $ where the sequences $(N_k)_{k \in \N}$ and $(X_{k,i})_{(k,i) \in \N^2}$ are i.i.d. and mutually independent, $N_1 \sim \cP(\lambda)$, and $X_{1,1} \sim Y_1$.

\begin{lem}[Link between discrete and continuous time]
\label{equivalent entre cas discret et continu}
With the formalism outlined above, if $X \sim Y$, then $\psi_c(u+c) \ \leq \ \psi_{dis}(u) \ \leq \ \psi_c(u)$.

In particular, if $\psi(u) \sim C.u^{-\gamma}$ when $u \to \infty$, with $C, \gamma >0$, then
$\psi_{dis}(u) \, \sim \, \psi(u)$.
\end{lem}

\begin{proof}
The key idea is that when $t=n$ is an integer, we have, regarding the notations, \\
$\sum\limits_{k=1}^n \sum\limits_{i=1}^{N_k} X_{k,i} \ \sim \ \sum\limits_{k=1}^{M(n)} Y_k \ .$
Then, we compare $(R_n)_{n \in \N}$ to $R_c(u,t)$ and $R_c(u+c,t)$: for a fixed $n \in \N^*$, and $t \in [n-1,n]$, we have $R_n \leq_{st} R_c(u+c,t)$. We deduce $\psi_c(u+c) \ \leq \ \psi_{dis}(u)$ since it is true for all integer $n$.

Otherwise, looking at the ruin of the process $(R_n)_{n \in \N}$ is equivalent, in terms of distributions, to looking at the ruin of $(R_c(u,t))_{t \in \R_+}$ only for the times where $t$ is an integer. It yields \\ 
$ \psi_{dis}(u) \leq \psi(u)$.
\end{proof}

This link between discrete and continuous time allows us to propose the following result:

\begin{prop}[Bound for a subexponential ruin probability]
\label{inégalité proba ruine sous-exp modèle 1D classique}
Let us consider the model from Equation \ref{modèle classique en 1D avec BC et INAR(1)}. We assume that $X$ is such that its stationary excess distribution $X_e$ is subexponential, and $\overline{G}(x) = O(\overline{F}(x))$, or conversely. \\ 
Then, we have $\psi(u) \ \geq \ \psi_{min} (u)$, with
\begin{equation}
\psi_{min}(u) \ \sim \ D_0  \int_u^\infty \left(\overline{F}(x) + \overline{G}(x) \right) \, dx \, , \ \,  \text{where} \ \, D_0 \ = \ \frac{\lambda}{c - \lambda(\mu_X + \mu_Y)} \  .
\end{equation}
\end{prop}

\begin{proof}
Since $N_k \leq M_k$ for each $k \geq 1$, by defining 
\begin{equation}
R^{min}_n \ = \ u \ + \ c \, n \ - \ \sum\limits_{k=1}^n \,  \sum\limits_{i=1}^{N_k}  (X_{k,i} + Y_{k,i}) \ \, ,
\end{equation}
we clearly have $R_n \, \leq \, R^{min}_n$ for each $n \in \N$. Then, the ruin probability $\psi_{min}(u)$ is deduced thanks to Equation \ref{equivalent continu loi sous-exp avec by-claims}, and Lemma \ref{equivalent entre cas discret et continu}. 
\end{proof}

\begin{cor}[Application for Pareto distributions] ~\\
\label{corollaire Pareto ruine ultime modèle classique 1D}
Let us consider the model from Equation \ref{modèle classique en 1D avec BC et INAR(1)}. We assume $X \sim \operatorname{Par}(\alpha,\theta)$, and $Y \sim \operatorname{Par}(\beta,\eta)$, with $\alpha, \beta >1$, et $\theta, \eta >0$. Then, regarding the framework from Proposition \ref{inégalité proba ruine sous-exp modèle 1D classique},
\begin{equation}
D_0 \ = \ \frac{\lambda  (\alpha -1) (\beta -1) }{c(\alpha-1)(\beta -1) - \lambda[\theta(\beta -1) + \eta(\alpha -1)]} \ .
\end{equation}

i) If $\alpha < \beta$, therefore $\psi_{min}(u) \ \sim \ D_0 \frac{\theta^\alpha}{\alpha - 1} \, u^{-\alpha + 1} \ .$
\smallskip

ii) If $\alpha > \beta$, then $\psi_{min}(u) \ \sim \ D_0  \frac{\eta^\beta}{\beta -1} \, u^{-\beta + 1} \ $.
\smallskip

iii) If $\alpha = \beta$, then
$\psi_{min}(u) \ \sim \ D_0  \frac{\theta^\alpha + \eta^\alpha}{\alpha - 1} \, u^{-\alpha + 1} \ .$
\end{cor}
\medskip

These results are true for the three models.

Actually, for the purely independent model, it is possible to go further and compute precisely the asymptotic ruin probability in such a context.

\begin{cor}[Subexponential asymptotic ruin probability] ~\\
\label{ruine ultime discrète avec 2 lois sous-exp}
Let us consider the model from Equation \ref{modèle purement indépendant en 1D}. We assume that $X$ is such that its stationary excess distribution $X_e$ is subexponential, and $\overline{G}(x) = O(\overline{F}(x))$, or conversely. \\ 
Then, when $u \to \infty$, the ultimate ruin probability satisfies:
\begin{equation}
\label{Ruine ultime du modèle 1 avec deux lois sous-exp}
\psi_I(u) \ \sim \ D_1  \int_u^\infty \left(\overline{F}(x) + \overline{G}\left((1-\rho)x)\right) \right) \, dx \, , \ \, \text{where} \ \, D_1 \ = \ \frac{\lambda(1-\rho)}{c(1-\rho) - \lambda[(1-\rho)\mu_X + \mu_Y]} \  .
\end{equation}
\end{cor}

\begin{rem}
Even if it is not always true, for most of the cases, if $X_e$ is subexponential, then $X$ is also subexponential.
\end{rem}

\begin{proof}
The proof for the converse case would be the same.

We are going to apply in particular Equation \ref{equivalent continu loi sous-exp avec by-claims} and Lemma \ref{equivalent entre cas discret et continu}.
\medskip

Let us fix the real numbers $\lambda$, $\gamma_0$, $c$ and $\rho$.

We define the i.i.d. sequences $(L_k)_{k \in \N}$, and for $q \in \N^*$, $(O^q_k)_{k \in \N}$, such that $O^q_1 \sim \cP(\gamma_q)$, and $L_1 \sim \cP \left( \frac{\lambda}{1-\rho} \right)$.
We assume that for a fixed $q \in \N^*$, $L$ and $O^q$ are independent, and for $q' \neq q$, the sequences $O^q$ and $O^{q'}$ are also independent. Moreover, each of these sequences is independent of the other sequences $N$, $M^I$, $X$, and $Y$.
\medskip

Let us start with the case where $(\gamma_k)_{k \in \N}$ is nonincreasing, which is equivalent to \\
$\gamma_0 \geq \frac{\lambda}{1-\rho}$. The converse would be similar.

Under the stochastic order, for a fixed $q \in \N^*$, and for $k \geq q$, \\
$L_k \, \leq_{st} \, M^I_k \, \leq_{st} \, O^q_k \, \leq_{st} \, O^1_k$.

For $q \in \N^*$, and $s \in \N^*$, we define for each $n \in \N^*$,
\begin{equation}
\left\{
\begin{array}{ll}
    P_n \ = \ u \ + \ c \, n \ - \ \sum\limits_{k=1}^n \,  \sum\limits_{i=1}^{N_k} X_{k,i} \ - \ \sum\limits_{k=1}^n \,  \sum\limits_{i=1}^{L_k} Y_{k,i}  \ ,
     &  \\ H^q_n \ = \ u \ + \ c \, n \ - \ \sum\limits_{k=1}^n \,  \sum\limits_{i=1}^{N_k} X_{k,i} \ - \ \sum\limits_{k=1}^n \,  \sum\limits_{i=1}^{O^q_k} Y_{k,i}  \ ,
     & \\ U^s_n \ = \ u \ + \ c \, n \ - \ \sum\limits_{k=1}^n \,  \sum\limits_{i=1}^{N_k} X_{k,i} \ - \ \sum\limits_{{k=s}}^n \,  \sum\limits_{i=1}^{O^s_k} Y_{k,i} \ - \ \sum\limits_{k=1}^{{s-1}} \sum\limits_{i=1}^{O^k_k} Y_{k,i} \ .
\end{array}
\right.
\end{equation}

Let us fix $n \in \N$.
We have clearly $U_n^s\leq_{st} R^I_n \leq_{st} P_n$ for each $s \in \N^*$. Moreover, $U_n^1 = H_n^1$, and for $q \in \N^*$ and $r \in \N$, we have $U_n^q \leq_{st} H_n^q$, but also $H_n^q \leq_{st} H_n^{q + r}$, and $U_n^q \leq_{st} U_n^{q + r}$ (see for example Theorems 1.A.3 and 1.A.4 from \cite{shaked2007stochastic}).

Actually, the processes $(U^s)_{s \in \N}$ will serve as intermediate steps to compare $R$ with the processes $(H^q)_{q \in \N}$.
\smallskip

The ultimate ruin probability of the process $(P_n)_{n \in N}$ satisfies, when $u \to \infty$, \\
$\psi_P(u) \sim D_1 \int_u^\infty \left(\overline{F}(x) \, + \, \overline{G}\left((1-\rho)x \right) \right) \, dx$, where $D_1$ has the expression outlined previously in Equation \ref{Ruine ultime du modèle 1 avec deux lois sous-exp}.
To justify it, we apply Equation \ref{equivalent continu loi sous-exp avec by-claims} in discrete time, which is possible due to Lemma \ref{equivalent entre cas discret et continu}. More precisely, the continuous-time model linked is
\begin{equation}
\label{equation modèle avec BC continu et fréquence modifiée}
P_t \ = \ u \, + \, c \, t \, - \, \sum\limits_{i=1}^{M(t)} \left( X_i + \frac{1}{1-\rho}Y_i \right) \quad \text{for} \ t \in \R_+ \ .
\end{equation}
where here, the i.i.d. variables $(X_i)_{i \in \N}$ has the same distribution as $X$, and the i.i.d variables $(Y_i)_{i \in \N}$ the same as $Y$ (and the two sequences are independent).

Actually, the comparison is appropriate since per "unit of time", the mean of the aggregate loss is the same for both processes. This unit of time, for $(P_n)_{n \in \N}$ is a period, whereas for $(P_t)_{t \in \R_+}$, it corresponds to the intervals of the form $]n,n+1]$, where $n \in \N^*$.
\smallskip

Similarly, it is possible to get the asymptotic ruin probability for $H^q$, by replacing the variable $\frac{Y_k}{1 - \rho} \, $ by $\, \frac{\gamma_q \, Y_k}{\lambda} \, . \ $ It yields 
$$ \psi_{H^q}(u) \sim \ D(q) \int_u^\infty \left(\overline{F}(x) \, + \, \overline{G}\left(\frac{\lambda}{\gamma_q} x \right) \right) \, dx \, , \ \ \text{where} \ \ D(q) = \frac{\lambda}{c - (\lambda \mu_X + \gamma_q \, \mu_Y)} \ .$$

For $U^2$, as we work with fixed parameters $\gamma_0$ and $\lambda$, there exists an integer $m \in \N^*$ such that $\gamma_2 \, m \geq \gamma_1$.
We already have $U^2_n(u) \leq_{st} H^2_n(u)$. Then, we note $S^{H^2}_n$ and $S^{U^2}_n$ the aggregate amounts of losses at time $n \in \N^*$ for $H^2$ and $U^2$.

Let us prove that $S_n^{H^2} \leq_{st} S_n^{U^2} \leq_{st} S_{n+m}^{H^2} \,$, which is equivalent to:

$$ \sum\limits_{k=1}^n \,  \sum\limits_{i=1}^{O^2_k} Y_{k,i} \ \ \leq_{st} \ \ \sum\limits_{k=2}^n \,  \sum\limits_{i=1}^{O^2_k} Y_{k,i} \ + \ \sum\limits_{i=1}^{O^1_1} Y_{1,i} \ \ \leq_{st} \ \ \sum\limits_{k=1}^{n+m} \,  \sum\limits_{i=1}^{O^2_k} Y_{k,i} \ .$$

For that, we define $V_n = \sum\limits_{k=1}^n O_k^2$, and $W_n = O_1^1 + \sum\limits_{k=2}^n O_k^2$, which are the respective total number of by-claims at time $n \in \N^*$ for $H^2$ and $U^2$. By independence, $V_n \sim \cP\left( n\gamma_2\right)$, and \\ 
$W_n \sim \cP\left( \gamma_1 + (n-1)\gamma_2\right)$. Thus, $V_n \leq_{st} W_n \leq_{st} V_{n+m}$.
\smallskip

As we are looking for results when $u \to \infty$, we can take $u$ such that $u \geq c \, m$.
Therefore, since $(Y_{k,i})_{(k,i) \in \N^2}$ is i.i.d., $H^2_{n+m}(u-cm)\  \leq_{st} \ U^2_n(u) \ \leq_{st} \ H^2_n(u)$.
\smallskip

Taking the ultimate ruin probabilities, $\psi_{H^2}(u-cm) \ \leq \ \psi_{U^2}(u) \ \leq \ \psi_{H^2}(u)$. Since \\
$\psi_{H^2}(u-cm) \, \sim \, \psi_{H^2}(u)$, we get
$\psi_{U^2}(u) \sim \frac{\lambda}{c - (\lambda \mu_X + \gamma_2 \, \mu_Y)}  \int_u^\infty \left(\overline{F}(x) \, + \, \overline{G}\left(\frac{\lambda}{\gamma_2} x \right)\right) \, dx$.
\medskip

Now, to prove what we want, we start an induction for $s \geq 2$, using the processes $(U_n^s)_{n \in \N}$.
We have already proved the case $s=2$. 

For $s \geq 3$, we have instead $V_n \sim \cP\left( n\gamma_s\right)$, and $W_n \sim \cP\left( \sum\limits_{i=1}^{s-1} \gamma_i + (n+1-s)\gamma_s\right)$. 
So we take $m \in \N^*$ such that $m \, \gamma_s \geq \sum\limits_{i=1}^{s-1} \gamma_i$.

Hence, $\psi_{U^{s}}(u) \, \sim \, \psi_{H^{s}}(u) \, \sim \, \frac{\lambda}{c - (\lambda \mu_X + \gamma_s \, \mu_Y)} \int_u^\infty \left(\overline{F}(x) \, + \, \overline{G}\left(\frac{\lambda}{\gamma_s} x \right) \right) \, dx$ when $u \to \infty$.

Thanks to the framework outlined above, we have
$\psi_P(u) \leq \psi_I(u) \leq \psi_{U^s}(u)$, for each $s \in \N^*$. Then, as $s \to \infty$, $\underset{s \to \infty}{\lim} \gamma_s \, = \, \frac{\lambda}{1 - \rho}$. 

All in all, it yields $\psi_I(u) \sim D_1  \int_u^\infty \left(\overline{F}(x)\, + \, \overline{G}\left((1-\rho) x \right) \right) \, dx \ .$
\medskip

If $(\gamma_k)_{k \in \N}$ is increasing, instead, for $k \geq q$, $O^q_k \leq_{st} M_k \leq_{st} L_k$, and $O^q_1 \leq_{st} O^{q'}_1$ for some integers $q \leq q'$. The rest would be similar, this time we obtain the framework \\
$\psi_{U^s}(u) \leq \psi_I(u) \leq \psi_{P}(u)$, but with the same asymptotic results.
\end{proof}
\medskip

Taking two Pareto distributions, we can apply the previous result:

\begin{cor}[Ultimate ruin probabilities with 2 Pareto distributions] ~\\
\label{1D Pareto ruine ultime}
Let us consider the model from Equation \ref{modèle purement indépendant en 1D}. We assume $X \sim \operatorname{Par}(\alpha,\theta)$, and $Y \sim \operatorname{Par}(\beta,\eta)$, with $\alpha, \beta >1$, and $\theta, \eta >0$. From that, we get:
$$D_1 \ = \ \frac{\lambda  (1 - \rho) (\alpha -1) (\beta -1) }{c(1-\rho)(\alpha-1)(\beta -1) - \lambda[(1-\rho)\theta(\beta -1) + \eta(\alpha -1)]} \ .$$
\medskip

i) If $\alpha < \beta$, therefore $\psi_I(u) \ \sim \ D_1 \frac{\theta^\alpha}{\alpha - 1} \, u^{-\alpha + 1} \ .$
\smallskip

ii) If $\alpha > \beta$, then $\psi_I(u) \ \sim \ D_1  \frac{\eta^\beta}{(1-\rho)^\beta(\beta -1)} \, u^{-\beta + 1} \ $.
\smallskip

iii) If $\alpha = \beta$, then
$\psi_I(u) \ \sim \ D_1  \frac{\theta^\alpha . (1 - \rho)^\alpha + \eta^\alpha}{(1 - \rho)^\alpha (\alpha - 1)} \, u^{-\alpha + 1} \ .$
\end{cor}
\bigskip

Then, we can focus on a particular class of subexponential distribution, the class of extended regularly varying (ERV) functions, which generalizes the class of regularly varying functions.
Other results involving the ERV class for models built with by-claims can be found in \cite{li2013pairwise} or \cite{fu2016asymptotic}.

The interest of such a kind of distribution is to provide a larger framework (even if the results are consequently less refined), in case there is uncertainty on the heaviness of the risks: typically, if the distribution of the main claims or the by-claims is assumed to be Pareto distributed, but the shape parameter is not known precisely.

\begin{defi}[ERV class]
Let $0 < \alpha \leq \beta$ be real numbers. A distribution \\
$F \in \operatorname{ERV}(- \alpha, - \beta)$, if for all $y \geq 1 \,$, 
$\ y^{- \beta } \ \leq \  \underset{x \to \infty}{\liminf} \frac{\overline{F}(xy)}{\overline{F}(x)} \ \leq \ \underset{x \to \infty}{\limsup} \frac{\overline{F}(xy)}{\overline{F}(x)} \ \leq \ y^{- \alpha} \, .$
The case $\alpha = \beta$ simply corresponds to the class $\mathcal{R}_{- \alpha}$.
\end{defi}

\begin{prop}[Finite-time for the ERV class] ~\\
\label{ERV en temps fini}
Let us consider the original model from Equation \ref{modèle classique en 1D avec BC et INAR(1)}.
We assume $F \in \operatorname{ERV}(-\alpha,-\beta)$, and \\
$G \in \operatorname{ERV}(-\delta,-\eta)$, with $1 < \alpha \leq \beta$ and $1 < \delta \leq \eta$. Then, for $T \in \N^*$,\\
i) If $\beta < \delta$, then $\psi_T(u) \in \operatorname{ERV}(-\alpha,-\beta)$ seen as a function of $u$, \\
ii) If $\alpha > \eta$, we get $\psi_T(u) \in \operatorname{ERV}(-\delta,-\eta)$, \\
iii) Otherwise,  $ \psi_T(u) \in \operatorname{ERV}(- (\alpha \wedge \delta), \, -( \beta \vee \eta ) ) $.
\end{prop}

\begin{proof}
The ERV distributions are subexponential, so with Lemma X.2.2 of \cite{asmussen2010ruin}, for $k \in \N^*$, when $x \to \infty$, $\overline{F_{A_k}}(x) \sim \lambda \overline{F}(x)$ and $\overline{F_{B_k}}(x) \sim \gamma_k \overline{G}(x)$. Then, the rest is similar as for Proposition \ref{1Ddeuxqueueslourdes}, and we distinguish the cases:
\medskip

The case i) implies $\overline{G}(x) = o ( \overline{F}(x))$, whereas for ii), $\overline{F}(x) = o (\overline{G}(x))$. 
\medskip

For iii), the key element is that if $F \in \operatorname{ERV}(- \alpha, - \beta)$, consequently, for $1 < a \leq \alpha$, and $\beta \leq b$, we have
$ \  y^{- b} \, \leq \, \underset{x \to \infty}{\liminf} \frac{\overline{F}(xy)}{\overline{F}(x)} \, \leq \, \underset{x \to \infty}{\limsup} \frac{\overline{F}(xy)}{\overline{F}(x)} \, \leq \, y^{- a} \ .$

Thus, $F \in \operatorname{ERV}(- (\alpha \wedge \delta), \, -( \beta \vee \eta ) \, )$ and $G \in \operatorname{ERV}(- (\alpha \wedge \delta), \, -( \beta \vee \eta ) \, )$. 
\end{proof}

This result could be extended to the two models with softened dependence.
\medskip

Now, in infinite time, the following result is only true for the purely independent model:

\begin{prop}[Ultimate ruin probability for the ERV class]~\\
\label{ERV en temps infini}
Let us consider the model from Equation \ref{modèle purement indépendant en 1D}. \\
Under the same assumptions, we look at the ruin probabilities under an infinite-time horizon. We draw the framework $K_1(u) \leq \psi_I(u) \leq K_2(u)$, where: \\
i) If $\beta < \delta$, then $\ K_1, K_2 \in \operatorname{ERV}(-\alpha+1,-\beta+1)$. \\
ii) If $\alpha > \eta$, we get $K_1, K_2 \in \operatorname{ERV}(-\delta+1, -\eta+1)$. \\
iii) Otherwise,  $K_1, K_2 \in \operatorname{ERV}(- (\alpha \wedge \delta)+1, \, -( \beta \vee \eta )+1 \, ) $.
\end{prop}

\begin{proof}
We deal only with the case i), because the two others would be similar.
\medskip

We set, for $(k,i) \in \N^2$, $Z_{k,i} = X_{k,i} + Y_{k,i}$. Thus, $(Z_{k,i})_{(k,i) \in \N^2}$ is a sequence of i.i.d. variables.
We also set a sequence $(L_k)_{k \in \N}$ of i.i.d. variables such that $L_1 \sim \cP(\gamma_{max})$. We assume that $L$ is independent of $N$, $M^I$, $X$ and $Y$. Then, we define
\begin{equation}
\left\{
\begin{array}{ll}
     R_n^{max} \ = \ u + c \, n \, - \, \sum\limits_{k=1}^n \, \sum\limits_{i=1}^{L_k} Z_{k,i} \quad ,
     & \\ R_n^{min} \ = \ u + c \, n \, - \, \sum\limits_{k=1}^n \,  \sum\limits_{i=1}^{N_k} Z_{k,i} \ .
\end{array}
\right.
\end{equation}
Clearly, $R_n^{max} \leq_{st} R^I_n \leq_{st} R_n^{min}$.

To do it properly, we define the continuous-time transcription of the processes: \\
$R^{min}_c(t) \ = \ u \, + \, ct \, - \, \sum\limits_{i=1}^{N(t)} Z_i$, and the same thing for $R^{max}_c$, replacing $N(t)$ by $L(t)$, where $(N(t))_{t \in \R_+} \sim \cP o(\lambda)$ and $(L(t))_{t \in \R_+} \sim \cP o(\gamma_{max})$ are Poisson processes.

Since $Z$ is subexponential, we apply Equation \ref{equivalent continu loi sous-exp}:
$\psi_c^{min}(u) \sim \frac{\lambda \, \mu_Z}{c - \lambda \,\mu_Z}  \int_u^\infty \overline{F_Z}(s) \, ds$, and $\psi_c^{max}(u) \sim \frac{\gamma_{max} \, \mu_Z}{c - \gamma_{max} \, \mu_Z}  \int_u^\infty \overline{F_Z}(s) \, ds \, $.

We note $H_Z(u) = \int_u^\infty \overline{F_Z}(s) \, ds$, and the same thing for $H_X$ and $H_Y$.
\medskip

Let $y \geq 1$ and $0 < \e < 1$ be real numbers.

By definition,
$\, \underset{x \to \infty}{\liminf} \, \frac{\overline{F}(xy)}{\overline{F}(x)} = \underset{x \to \infty}{\lim} \left[ \inf \left( \frac{\overline{F}(zy)}{\overline{F}(z)}, \, z \geq x \right) \right] \, $. \\
Since $X \in \erv (-\alpha, - \beta)$, it means there exists $M >0$ such that for each real number $x \geq M$, we have $\frac{\overline{F}(xy)}{\overline{F}(x) \, y^{- \beta}} \ \geq \ 1 - \e$. \ 
It yields $\, \overline{F}(xy) \, \geq \, (1 - \e) \overline{F}(x)  y^{-\beta}$.

In particular, for $u \geq M$, $\int_u^\infty \overline{F}(sy) \, ds \geq (1-\e)y^{-\beta} \int_u^\infty \overline{F}(s) \, ds$, which leads to \\
$\ y^{-1} \int_{uy}^\infty \overline{F}(s) \, ds \geq (1-\e)y^{-\beta} \int_u^\infty \overline{F}(s) \, ds \, , \ $ so
$\, \frac{H_X(uy)}{H_X(u)} \ \geq \ (1-\e) y^{- \beta + 1}$.

Since it is true for every $\e >0$, we deduce that $\ \underset{u \to \infty}{\liminf} \frac{H_X(uy)}{H_X(u)} \geq y^{- \beta + 1} $.

Similarly, we may prove that $\, \underset{u \to \infty}{\limsup} \frac{H_X(uy)}{H_X(u)} \leq y^{- \alpha + 1} $. \\
Thus, $H_X \in \erv(-\beta+1,-\alpha +1)$.
\medskip

Moreover, since $\beta < \delta $, $\, \overline{G}(x) = o(\overline{F}(x))$. Therefore, $\overline{F_Z}(x) \sim \overline{F}(x)$.

What is more, the integral $\int_u^\infty \overline{F_Z}(s) \, ds$ converges for every $u \geq 0$, so thanks to asymptotic results for integration, when $u \to \infty$, we have $H_X(u) \sim H_Z(u)$.

We deduce easily that, for $y \geq 1$, when $u \to \infty$, $\frac{H_Z(yu)}{H_Z(u)} \sim \frac{H_X(yu)}{H_X(u)}$, and then \\ $H_Z \in \erv(-\beta+1,-\alpha +1)$.
\medskip

Therefore, $\psi_c^{min} (u)  \in \erv(-\beta+1,-\alpha +1)$, and same thing for $\psi_c^{max}(u)$. Moreover, thanks to Lemma \ref{equivalent entre cas discret et continu}, we get
\begin{equation}
\psi_c^{min} (u+c) \leq \psi^{min} (u) \leq \psi_I(u) \leq \psi^{max}(u) \leq \psi_c^{max} (u) \ .     
\end{equation}
Since $H_Z$ belongs to the ERV class, it is subexponential and thus long-tailed. Hence, when \\
$u \to \infty$, $H_Z(u) \sim H_Z(u+c)$, and so $\psi_c^{min} (u) \sim \psi_c^{min} (u+c)$.

Thanks to this framework, we eventually obtain what we wanted.
\end{proof}

For the original or the half-independent models, because of the corresponding dependence assumptions, it is not possible to establish the previous upper bound (or at least, to derive it the same way).

\begin{rem}
\label{R - alpha est troué}
In general, the framework $f_1(u) \leq G(u) \leq f_2(u)$, with $f_1, f_2 \in \cR_{-\alpha}$ provides an idea of the asymptotic behaviour of the function $G$, but it does not ensure that $G \in \cR_{-\alpha}$. Of course, this remains true if $f_1, f_2 \in \operatorname{ERV}(- \alpha, - \beta)$ instead.
\end{rem}

What is more, we summarise the possible extensions of these results with heavy-tailed distributions:

\begin{rem}
Regarding Corollary \ref{1D Pareto ruine ultime}, and Propositions \ref{1Ddeuxqueueslourdes}, \ref{ERV en temps fini}, or \ref{ERV en temps infini}, we would get the same kind of results by exchanging some of the assumptions, as long as we keep the idea of dominance of a given tail by a heavy-tailed distribution.

For instance, we may replace the distribution of $Y$ by an exponential or a light-tailed distribution, or conversely towards $X$.
Otherwise, it is possible to mix the heavy-tailed distributions, for example a Pareto one with an ERV one, etc.
\end{rem}

\subsection{The BRV class, a particular case of dependence}
\label{section BRV}

It is possible to relax the independence assumptions of our model, and focus on the specific case of the bivariate regular variations class: here, the sequences representing the amounts of the losses $(X_{k,i})_{(k,i) \in \N^2}$ and $(Y_{k,i})_{(k,i) \in \N^2}$, even if the bidimensional vectors $(X_{k,i}, Y_{k,i})_{(k,i) \in \N^2}$ are still assumed to be independent.

Actually, as a practical point of view, the amount of a main claim should not be independent of the amounts of the associated by-claims. As a simple example, regarding the PFAS issue, the more molecules are released through emissions, the more it could cause damages towards the environment or human health.

\begin{defi}{(Bivariate regular variation)}
A random pair $(X,Y)$ is said to follow a bivariate regularly varying distribution if there exists a distribution $F_0 \in \mathcal{R}_{-\alpha}$ and a
nondegenerate limit measure $\nu$ on $[0,\infty]^2 \, \backslash  \{0\} $, namely $(X,Y) \in \operatorname{BRV}_{- \alpha}(\nu, F_0)$, if for every Borel set $B$ $\nu$-continuous (i.e. such that $\nu (\partial B) = 0$), we have
$$
\frac{1}{\overline{ F_0} (t)} \p \left( \frac{1}{t} (X,Y) \in B \right) \underset{t \to \infty}{\longrightarrow} \nu(B) \ .
$$
It corresponds to a multivariate heavy-tailed distribution.
\end{defi}

\begin{prop}{(Finite-time with a BRV pair)} ~\\
Let us consider the original model from Equation \ref{modèle classique en 1D avec BC et INAR(1)}. \\
Let us set $A = \{ (x,y) \in \R^2 \ / \ x + y > 1 \}$.
We assume there exists $F_0 \in \cR_{-\alpha}$, ($\alpha >1$ to fulfill the net profit), and a nondegenerate measure $\nu$ on $[0,\infty]^2 \, \backslash  \{0\} $ such that $\nu(A) > 0$, $A$ is a $\nu$-continuous set, and
$(X_{1,1}, Y_{1,1}) \in \operatorname{BRV}_{-\alpha}(\nu,F_0)$. \\
Let us fix $T \in \N^*$. Then, there exists $H_{1,T}, H_{2,T} \in \cR_{-\alpha}$ such that
\begin{equation}
H_{1,T}(u) \ \leq \ \psi_T(u) \ \leq \ H_{2,T}(u) \ .
\end{equation}
\end{prop}

\begin{proof}
We pick up some elements of the proof of Proposition \ref{ERV en temps infini}. For $n \in \N$, we keep the same expression for $R^{min}_n$. For $R^{max}_n$, we can be more precise here:
$$ R^{max}_n \ = \ u \, + \, c \, n \, - \,  \sum\limits_{k=1}^n \left( \sum\limits_{i=1}^{M_k} (X_{k,i} + Y_{k,i}) \right) \ . $$
Directly, $R_n^{max} \ \leq \ R_n \ \leq \ R_n^{min} \, .$ \\
Then, if $(X, Y) \in \operatorname{BRV}_{-\alpha}(\nu,F_0)$, for $u >0$,
$\p(X + Y > u) = \p \left( \frac{X +  Y}{u} > 1 \right)$, so \\
$\p(X + Y > u)  = \p \left( \frac{(X, Y)}{u} \in A \right)$,
which yields $\p(X + Y > u) \sim \nu(A) \, \overline{F_0} (u)$ when $u \to \infty$.
\smallskip

Eventually, we operate successive convolutions (see e.g. Proposition X.1.9 of \cite{asmussen2010ruin}), and we obtain the framework outlined above, with
\begin{equation}
\label{équation d'encadrement BRV temps fini}
\left\{
\begin{array}{ll}
        H_{1,T}(u) \, = \, \psi_T^{min}(u) \, \sim \, T  \lambda \, \nu(A) \, \overline{F_0}(u)  \, ,
     & \\ H_{2,T}(u) \, = \, \psi_T^{max}(u) \, \sim \, \left(\sum\limits_{k=1}^T \gamma_k \right) \nu(A) \,\overline{F_0}(u) \, .
\end{array}
\right.
\end{equation}
\end{proof}

This result could be extended to the two models with softened dependence.

Once again, as mentioned in Remark \ref{R - alpha est troué}, this kind of framework is not enough to deduce that $\psi_T(u)$ is $\cR_{-\alpha}$, even if it suggests an idea of its asymptotic behaviour.
\medskip

With such a framework, it is also possible to propose an ultimate ruin probability for the purely independent model (but not for the other models):

\begin{prop}{(Ultimate ruin probability for a BRV pair)} ~\\
\label{BRV en temps infini}
Let us consider the model from Equation \ref{modèle purement indépendant en 1D}.
Under the same assumptions, we focus on an infinite-time horizon. \\
There exists $H_1, H_2 \in \cR_{-(\alpha -1)}$ such that $H_1(u) \ \leq \ \psi_I(u) \ \leq \ H_2(u) \ .$
\end{prop}

\begin{proof}
We keep the notations of Proposition \ref{ERV en temps infini}. In particular, when $u \to \infty$, \\ $\overline{F_Z}(u) = \p(X_{1,1} + Y_{1,1} > u) \sim \nu(A) \overline{F_0}(u)$, and since $F_0 \in \cR_{-\alpha}$, we deduce $F_Z \in \cR_{-\alpha}$.

Thus, $F_Z$ is a subexponential distribution, we can apply Equation \ref{equivalent continu loi sous-exp} and Lemma \ref{equivalent entre cas discret et continu}, to deduce that $\psi^{min}(u) \sim \frac{\lambda}{c - \lambda \mu_Z} \, \int_u^\infty \overline{F_Z}(y) \, dy$, with $\mu_Z = \mu_X + \mu_Y$. 

Thanks to 
Karamata's theorem (see e.g. \cite{bingham1989regular}), there exists a slowly varying function $L$ such that
$\int_u^\infty \overline{F_Z}(y) \, dy \sim \frac{\nu(A) \, L(u)}{(\alpha -1) \, u^{\alpha-1}}$.
\bigskip

For the upper bound, we derive the asymptotic ruin probability $\psi^{max}(u)$ related to $R^{max}$ similarly, and 
we eventually deduce the framework outlined:
\begin{equation}
\label{équation d'encadrement BRV temps infini}
\left\{
\begin{array}{ll}
       \vspace{0.3 cm}  H_1(u) \ = \ \psi^{min}(u) \, \sim \, \frac{\lambda \, \nu(A) L(u)}{[c - \lambda (\mu_X + \mu_Y)](\alpha - 1)} \, u^{-\alpha + 1}\, \ ,
     & \\ H_2(u) \ = \ \psi^{max}(u) \, \sim \, \frac{\gamma_{max} \, \nu(A) L(u)}{[c - \gamma_{max} (\mu_X + \mu_Y)](\alpha - 1)} \, u^{-\alpha + 1} \, .
\end{array}
\right.
\end{equation}
\end{proof}

In fact, it is not possible to derive the upper bound in the previous result for the two other models (or at least, to prove it the same way).
\medskip

A similar result is found for instance in \cite{li2017note}, or \cite{yang2019asymptotic}, describing continuous-time models where one main claim engenders exactly one by-claim (which may occur after a random bounded delay). 

With the same kind of assumptions for the pair $(X,Y)$, it is found that \\
$\psi(u) \sim \frac{\lambda \, \nu(A)}{c-\lambda(\mu_X + \mu_Y)} \int_{u}^{\infty} \overline{F_0}(y) dy \ ,$ which actually corresponds to the first result of Equation \ref{équation d'encadrement BRV temps infini}, but written differently.
\medskip

Moreover, we can state that in both finite and infinite time, the asymptotic power of the ruin probability is the same as if the two distributions of the losses were independent, with one from the class $\cR_{-\alpha}$ dominating the other (as in Proposition \ref{1Ddeuxqueueslourdes} and Corollary \ref{1D Pareto ruine ultime}).

\section{Numerical results}
\label{section résultats numériques article 1D}

In this section, we yield numerical results related with the model defined by Equation \ref{modèle classique en 1D avec BC et INAR(1)}, through some different approaches. First, we provide some ruin probabilities. Then, we study the influence of the parameters $\rho$ and $\lambda$, and what could happen if we underestimate the value of one of these parameters. Eventually, closely linked to Solvency II, we try to draw predictions and compute a ten-period ruin probability for some set of parameters satisfying $\psi_1(u) \simeq 0.005$. 

Some technical precisions: for the Pareto distribution, we keep Definition \ref{définition de la loi de Pareto}. Moreover, a confidence level of 99\% is applied for every margin displayed.

Eventually, to fit with the theoretical literature (see for instance, \cite{asmussen2010ruin}), we choose each time a realistic safety loading, i.e. $12.5 \% \leq \chi \leq 20 \%$. Specifically, we fix each time $\lambda \geq \gamma_0$, which means $\gamma_{max} = \frac{\lambda}{1 - \rho}$, and therefore,
\begin{equation}
\label{équation de chargement de sécurité pour les simulations}
\chi \ = \ \frac{\lambda(\mu_X(1-\rho) + \mu_Y)}{c(1-\rho)} \ - \ 1  \ .  
\end{equation}

\subsection{First sets of ruin probabilities}

\subsubsection{Starting with a small value of $\rho$}

In this paragraph, we fix $\lambda = 10$, $\gamma_0 = 10$, and $\rho = 0.1$, and we are observing $T=10$ periods.

We set $c = 37$, which leads to a safety loading $\chi \simeq 18.9\%$. 

We vary the distributions of $X$ and $Y$, and we give the ruin probabilities as functions of $u$. We study four cases: one with two heavy-tailed (Pareto) distributions, one with two light-tailed (exponential) distributions, and two cases with one Pareto distribution and one exponential distribution, mixing this way an heavy-tailed and a light-tailed distribution.

\begin{table}[H]
\centering
\resizebox{0.8\textwidth}{!}{
\newcolumntype{C}{>{\centering\arraybackslash}X}
\begin{tabularx}{\textwidth}{|C|C|C|C|C|}
\toprule
& \multicolumn{2}{|c|}{$X \sim \Exp(1/2)$, \, $Y \sim \Exp(1)$} & \multicolumn{2}{|c|}{$X \sim \Exp(1/2)$, \, $Y \sim \Par(3,2)$}  \\
\midrule
$u$  & $\psi_{10}(u)$ & Confidence & $\psi_{10}(u)$ & Confidence  \\
\midrule
0 & 0.452 & 0.00128 & 0.465 & 0.00129 \\
\midrule
1 & 0.427 & 0.00128  & 0.442 & 0.00128  \\
\midrule
5 & 0.337 & 0.00121 & 0.357 & 0.00124  \\
\midrule
10 & 0.244 & 0.00111 & 0.268 & 0.00114   \\
\midrule
20 & 0.121 & 0.00089 & 0.147 & 0.00091 \\
\midrule
30 & 0.058 &  0.00058 & 0.079 & 0.000694  \\
\midrule
50 & 0.012 & 0.00031 & 0.022 & 0.00038 \\
\midrule
100 & 0.00023 & 0.000074  & 0.0013 & 0.000093  \\
\midrule
\end{tabularx}
}
\caption{Ruin probabilities with $X \sim \Exp(1/2)$ for $\rho = 0.1$.
\label{Tableau de ruine avec rho = 0.1, et X de loi Exp}}
\end{table}

\begin{table}[H]
\centering
\resizebox{0.8\textwidth}{!}{
\newcolumntype{C}{>{\centering\arraybackslash}X}
\begin{tabularx}{\textwidth}{|C|C|C|C|C|}
\toprule
& \multicolumn{2}{|c|}{$X \sim \Par(2,2)$, \, $Y \sim \Exp(1)$} & \multicolumn{2}{|c|}{$X \sim \Par(2,2)$, \, $Y \sim \Par(3,2)$}  \\
\midrule
$u$  & $\psi_{10}(u)$ & Confidence & $\psi_{10}(u)$ & Confidence  \\
\midrule
0 & 0.476 &  0.00129  & 0.489 &  0.00129 \\
\midrule
1 & 0.458 & 0.00129 & 0.470 & 0.00129 \\
\midrule
5 & 0.396 & 0.00126 & 0.411 & 0.00127  \\
\midrule
10 & 0.332 & 0.00122  & 0.346 & 0.00122 \\
\midrule
25 & 0.201 & 0.00104  & 0.214 &  0.00106 \\
\midrule
50 & 0.01 & 0.00077  & 0.108 &  0.00080  \\
\midrule
100 & 0.034 & 0.00047 & 0.036 & 0.00048 \\
\midrule
200 & 0.009 & 0.00024 & 0.0096 & 0.00025  \\
\midrule
500 & 0.0015 & 0.000010 & 0.0015 & 0.00010 \\
\midrule
\end{tabularx}
}
\caption{Ruin probabilities with $X \sim \Par(2,2)$ for $\rho = 0.1$.
\label{Tableau de ruine avec rho = 0.1, et X de Pareto}}
\end{table}

We state that the results are quite similar when
$X \sim \Exp(1/2)$ on the one hand, or when \\ 
$X \sim \Par(2,2)$ on the other hand, but each time they are a little more risky when $Y \sim \Par(3,2)$.

\subsubsection{Increasing the value of the reproduction coefficient}
We almost keep the same safety loading (here, $\chi = 20\%$, contrary to $\chi = 18.9 \%$ before).

We increase the value of $\rho$, and display the results: $\rho=0.4$ (combined with $c=44$), and $\rho=0.7$ (combined with $c=64$). Otherwise, we keep the same values for the other parameters.

However, we do not display all the cases.
\newpage

For $\rho = 0.4$, when $X \sim \Exp(1/2)$:

\begin{table}[H]
\centering
\resizebox{0.8\textwidth}{!}{
\newcolumntype{C}{>{\centering\arraybackslash}X}
\begin{tabularx}{\textwidth}{|C|C|C|C|C|}
\toprule
& \multicolumn{2}{|c|}{$X \sim \Exp(1/2)$, \, $Y \sim \Exp(1)$} & \multicolumn{2}{|c|}{$X \sim \Exp(1/2)$, \, $Y \sim \Par(3,2)$}  \\
\midrule
$u$  & $\psi_{10}(u)$ & Confidence & $\psi_{10}(u)$ & Confidence  \\
\midrule
0 & 0.328 & 0.00383 & 0.353 & 0.0039 \\
\midrule
1 & 0.306 & 0.00376 & 0.331 & 0.00384 \\
\midrule
5 & 0.234 & 0.00345 & 0.266 & 0.0036  \\
\midrule
10 & 0.166 & 0.00304 & 0.199 & 0.00326    \\
\midrule
20 & 0.081 & 0.00223 & 0.111 & 0.00256  \\
\midrule
30 & 0.04 & 0.0016 & 0.061 & 0.00195   \\
\midrule
50 & 0.00895 & 0.00077 & 0.0195 & 0.00113   \\
\midrule
100 & 0.00025 & 0.00013 & 0.0015 & 0.00031   \\
\midrule
\end{tabularx}
}
\caption{Ruin probabilities with $X \sim \Exp(1/2)$ for $\rho = 0.4$.
\label{Tableau de ruine avec rho = 0.4, et X de loi Exp}}
\end{table}
\medskip

For $\rho = 0.7$, when $X \sim \Par(2,2)$:

\begin{table}[H]
\centering
\resizebox{0.8\textwidth}{!}{
\newcolumntype{C}{>{\centering\arraybackslash}X}
\begin{tabularx}{\textwidth}{|C|C|C|C|C|}
\toprule
& \multicolumn{2}{|c|}{$X \sim \Par(2,2)$, \, $Y \sim \Exp(1)$} & \multicolumn{2}{|c|}{$X \sim \Par(2,2)$, \, $Y \sim \Par(3,2)$}  \\
\midrule
$u$  & $\psi_{10}(u)$ & Confidence & $\psi_{10}(u)$ & Confidence  \\
\midrule
0 & 0.125 & 0.0027 & 0.141 & 0.00284 \\
\midrule
1 & 0.12 & 0.00265 & 0.136 & 0.00279 \\
\midrule
5 & 0.104 & 0.00249 & 0.118 & 0.00263  \\
\midrule
10 & 0.087 & 0.0023 & 0.098 & 0.00243    \\
\midrule
25 & 0.056 & 0.00187 & 0.064 & 0.002  \\
\midrule
50 & 0.032 & 0.00144 & 0.035 & 0.00151   \\
\midrule
100 & 0.014 & 0.00097 & 0.016 & 0.00102   \\
\midrule
200 & 0.0055 & 0.00060 & 0.0053 & 0.00059   \\
\midrule
500 & 0.00111 & 0.00027 & 0.00115 & 0.00028   \\
\midrule
\end{tabularx}
}
\caption{Ruin probabilities with $X \sim \Par(2,2)$ for $\rho = 0.7$.
\label{Tableau de ruine avec rho = 0.7, et X de Pareto}}
\end{table}

Regarding all the tables for $\rho = 0.1$, $\rho = 0.4$ and $\rho = 0.7$, it is interesting to see that for a same safety loading, a same set of values for the parameters $(u,\lambda, \gamma_0)$, and the same distributions for $X$ and $Y$, the ruin probability $\psi_{10}(u)$ decreases in function of $\rho$. It looks quite counterintuitive, but it might due to the fact that here, $\mu_X$ and $\mu_Y$ have the same order of magnitude.
In particular, it seems to decrease faster when $X$ is exponential (than when it is Pareto distributed). 

However, if we exchange the distributions of $X$ and $Y$, we outline the reversed assertions.

\subsection{Ignoring some effects}

All the results shown here are related with quite good confidence, the margin $M$ is below $0.005$. When the results become particularly close to 0 or 1 (typically $\psi \leq 0.01$), it becomes a bit less precise, with the quotient $\psi/M \,$ smaller than 15 or 10 sometimes, but it doesn't really matter since it corresponds to cases where the ruin is close to be almost sure or almost impossible.

In this subsection, we are only focusing on cases with two heavy-tailed (Pareto) distributions or two light-tailed (exponential) distributions, but we vary the means $\mu_X$ and $\mu_Y$.

Thus, in each table, the case 1 corresponds to $X \sim \Par(2,2)$ and $Y \sim \Par(3,2)$, the case 2 to $X \sim \Exp(1/2)$ and $Y \sim \Exp(1)$, the case 3 to $X \sim \Par(3,2)$ and $Y \sim \Par(2,2)$, the case 4 to $X \sim \Exp(1)$ and $Y \sim \Exp(1/2)$.

\subsubsection{Ignoring or underestimating the contagion effects}

We would like to explore in this paragraph what could be the consequences of understating or even ignoring the contagion effects. 

Specifically, we take the example of an insurance company which determines the premium amounts without considering the potential contagion, i.e. which considers that $\rho = 0$ regarding Equation \ref{modèle classique en 1D avec BC et INAR(1)}. Then, we increase $\rho$, and we look at the consequences on the risk in terms of ruin probability.
Actually, these results should also represent quite well what should happen when $\rho$ is simply underestimated and not just ignored.

In that context, the safety loading becomes $\chi = \frac{c}{\lambda(\mu_X + \mu_Y)} - 1$.
\bigskip

We fix $\lambda = \gamma_0 = 10$. We assume $\mu_X + \mu_Y = 3$, more precisely with $\mu_X , \mu_Y \in \{ 1, 2\}$. To ensure a safety loading of $\chi = 20 \%$, we set $c= 36$.

It is interesting to determine, in function of $\mu_X$ and $\mu_Y$, the threshold value over which the safety loading $\chi$ becomes negative. Actually, $\chi(\rho) = 0 \iff \rho = \rhm$.

For the first two cases, $\mu_X = 2$, and $\mu_Y = 1$, and $\rhm = 0.375$.

For the last two cases, $\mu_X = 1$, and $\mu_Y = 2$, and $\rhm \simeq 0.2308$.
\medskip

We first display the results with a null initial reserve, $u =0$, and then when $u=100$.

\begin{table}[H]
\centering
\resizebox{0.75\textwidth}{!}{
\newcolumntype{C}{>{\centering\arraybackslash}X}
\begin{tabularx}{\textwidth}{|C|C|C|C|C|C|C|C|C|}
\toprule
 & \multicolumn{2}{|c|}{Case 1} & \multicolumn{2}{|c|}{Case 2} & \multicolumn{2}{|c|}{Case 3} & \multicolumn{2}{|c|}{Case 4} \\
\midrule
$\rho$  & $\psi_{1}(u)$ & $\psi_{10}(u)$ & $\psi_{1}(u)$ & $\psi_{10}(u)$ & $\psi_{1}(u)$ & $\psi_{10}(u)$ & $\psi_{1}(u)$ & $\psi_{10}(u)$  \\
\midrule
0 & 0.262 & 0.480 & 0.284 & 0.446 & 0.261 & 0.482 & 0.284 & 0.445 \\
\midrule
0.1 & 0.280 & 0.525 & 0.313 & 0.509 & 0.302 & 0.578 & 0.344 & 0.580  \\
\midrule
0.2 & 0.301 & 0.582 & 0.343 & 0.586 & 0.348 & 0.688 & 0.402 & 0.726  \\
\midrule
0.3 & 0.326 & 0.649 & 0.374 & 0.678 & 0.392 & 0.803 & 0.466 & 0.868   \\
\midrule
0.4 & 0.347 & 0.729 & 0.404 & 0.776 & 0.432 & 0.904 & 0.528 & 0.956  \\
\midrule
0.5 & 0.372 & 0.821 & 0.434 & 0.878 & 0.483 & 0.969 & 0.588 & 0.992  \\
\midrule
0.6 & 0.397 & 0.911 & 0.469 & 0.956 & 0.528 & 0.995 & 0.645 & 0.999  \\
\midrule
0.7 & 0.421 & 0.976 & 0.5 & 0.993 & 0.571 & 0.9998 & 0.696 & 1  \\
\midrule
0.8 & 0.449 & 0.998 & 0.535 & 1 & 0.613 & 1 & 0.747 & 1  \\
\midrule
0.9 & 0.477 & 1 & 0.566 & 1 & 0.659 & 1 & 0.791 & 1  \\
\midrule
\end{tabularx}
}
\caption{Underestimating $\rho$ for 4 different cases, with $u =0$. 
\label{Sous-estimer rho avec u=0}}
\end{table}
\bigskip

\begin{table}[H]
\centering
\resizebox{0.75\textwidth}{!}{
\newcolumntype{C}{>{\centering\arraybackslash}X}
\begin{tabularx}{\textwidth}{|C|C|C|C|C|C|C|C|C|}
\toprule
 & \multicolumn{2}{|c|}{Case 1} & \multicolumn{2}{|c|}{Case 2} & \multicolumn{2}{|c|}{Case 3} & \multicolumn{2}{|c|}{Case 4} \\
\midrule
$\rho$  & $\psi_{1}(u)$ & $\psi_{10}(u)$ & $\psi_{1}(u)$ & $\psi_{10}(u)$ & $\psi_{1}(u)$ & $\psi_{10}(u)$ & $\psi_{1}(u)$ & $\psi_{10}(u)$  \\
\midrule
0 & 0.0035 & 0.034 & 0 & 0.00013 & 0.0039 & 0.036 & 0 & 0.00016 \\
\midrule
0.1 & 0.0036 & 0.041 & 0 & 0.0003 & 0.0044 & 0.053 & 0 & 0.0014  \\
\midrule
0.2 & 0.0037 & 0.049 & 0 & 0.0012 & 0.0048 & 0.086 & 0 & 0.0105  \\
\midrule
0.3 & 0.0039 & 0.065 & 0 & 0.005 & 0.006 & 0.158 & 0 & 0.067   \\
\midrule
0.4 & 0.004 & 0.094 & 0 & 0.021 & 0.0064 & 0.303 & 0 & 0.266  \\
\midrule
0.5 & 0.0041 & 0.156 & 0 & 0.081 & 0.0067 & 0.558 & 0 & 0.630  \\
\midrule
0.6 & 0.0042 & 0.298 & 0 & 0.268 & 0.0076 & 0.834 & 0 & 0.921  \\
\midrule
0.7 & 0.0044 & 0.575 & 0 & 0.636 & 0.009 & 0.977 & 0 & 0.995  \\
\midrule
0.8 & 0.0047 & 0.886 & 0 & 0.938 & 0.0097 & 0.999 & 0 & 1  \\
\midrule
0.9 & 0.0049 & 0.995 & 0 & 0.998 & 0.0107 & 1 & 0 & 1  \\
\midrule
\end{tabularx}
}
\caption{Underestimating $\rho$ for 4 different cases, with $u =100$. 
\label{Sous-estimer rho avec u=100}}
\end{table}

As we could expect, the risk becomes high if $\rho$ is underestimated. However, with a comfortable initial reserve (i.e., $u=100$), the risk at a one-period horizon is very low, and even null for sometimes. However, for a ten-period horizon, such an initial reserve is not enough to prevent the risks incurred when $\rho$ is high (but if $\rho$ is quite low, then the risks remain reasonable).

\subsubsection{Underestimating the frequency}

We now look at the case where the frequency is underestimated.

For each case, we assume the safety loading is determined from the value $\lambda = 10$, then we observe what happens when $\lambda$ increases.

We set $\gamma_0 = 10$, and $u=0$.

Once again, it is interesting to determine, in function of $\mu_X$ and $\mu_Y$, the threshold value $\lambda_{max}$ over which the safety loading $\chi$ becomes negative, by solving $\chi(\lambda) = 0$.

With $\mu_X = 2$ and $\mu_Y =1$, we take $c = 37$, so $\chi \simeq 18.9 \%$, and $\lambda_{max} \simeq 11.89$.

With $\mu_X = 1$ and $\mu_Y =2$, we take $c = 38.5$, so $\chi \simeq 19.5 \%$, and $\lambda_{max} \simeq 11.94$.
\smallskip

\newpage
For $\rho = 0.1$, we display the following results.

\begin{table}[H] 
\centering
\resizebox{0.7\textwidth}{!}{
\newcolumntype{C}{>{\centering\arraybackslash}X}
\begin{tabularx}{\textwidth}{|C|C|C|C|C|C|C|C|C|}
\toprule
& \multicolumn{4}{|c|}{$c = 37$, and $\chi \simeq 18.9 \%$} & \multicolumn{4}{|c|}{$c = 38.5$, and $\chi \simeq 19.5 \%$} \\
\midrule
& \multicolumn{2}{|c|}{Case 1} & \multicolumn{2}{|c|}{Case 2} & \multicolumn{2}{|c|}{Case 3} & \multicolumn{2}{|c|}{Case 4} \\
\midrule
$\lambda$  & $\psi_{1}(u)$ & $\psi_{10}(u)$ & $\psi_{1}(u)$ & $\psi_{10}(u)$ & $\psi_{1}(u)$ & $\psi_{10}(u)$ & $\psi_{1}(u)$ & $\psi_{10}(u)$  \\
\midrule
10 & 0.262 & 0.489 & 0.285 & 0.452 & 0.256 & 0.476 & 0.277 & 0.436 \\
\midrule
11 & 0.327 & 0.624 & 0.375 & 0.645 & 0.318 & 0.615 & 0.363 & 0.625  \\
\midrule
12 & 0.396 & 0.757 & 0.47 & 0.817 & 0.383 & 0.746 & 0.453 & 0.804  \\
\midrule
13 & 0.466 & 0.866 & 0.555 & 0.928 & 0.45 & 0.855 & 0.541 & 0.920   \\
\midrule
15 & 0.602 & 0.976 & 0.714 & 0.996 & 0.588 & 0.971 & 0.702 & 0.995  \\
\midrule
17 & 0.723 & 0.998 & 0.832 & 0.9999 & 0.706 & 0.997 & 0.812 & 0.9999  \\
\midrule
20 & 0.858 & 1 & 0.936 & 1 & 0.844 & 1 & 0.928 & 1  \\
\midrule
25 & 0.963 & 1 & 0.99 & 1 & 0.957 & 1 & 0.989 & 1   \\
\midrule
30 & 0.994 & 1 & 0.999 & 1 & 0.991 & 1 & 0.999 & 1  \\
\midrule
40 & 0.9998 & 1 & 1 & 1 & 0.9999 & 1 & 1 & 1   \\
\midrule
50 & 1 & 1 & 1 & 1 & 1 & 1 & 1 & 1   \\
\midrule
\end{tabularx}
}
\caption{Underestimating $\lambda$ for 4 different cases, with $u=0$ and $\rho = 0.1$. 
\label{Sous-estimer rho avec u=0}}
\end{table}
\smallskip

For $\rho = 0.4$, we get the following results.
With $\mu_X = 2$ and $\mu_Y =1$, we take $c = 44$, so $\chi = 20 \%$, and $\lambda_{max} = 12$.
With $\mu_X = 1$ and $\mu_Y =2$, we take $c = 52$, so $\chi = 20 \%$, and $\lambda_{max} = 12$.

\begin{table}[H]
\centering
\resizebox{0.7\textwidth}{!}{
\newcolumntype{C}{>{\centering\arraybackslash}X}
\begin{tabularx}{\textwidth}{|C|C|C|C|C|C|C|C|C|}
\toprule
& \multicolumn{4}{|c|}{$c = 44$, and $\chi = 20 \%$} & \multicolumn{4}{|c|}{$c = 52$, and $\chi = 20 \%$} \\
\midrule
& \multicolumn{2}{|c|}{Case 1} & \multicolumn{2}{|c|}{Case 2} & \multicolumn{2}{|c|}{Case 3} & \multicolumn{2}{|c|}{Case 4} \\
\midrule
$\lambda$  & $\psi_{1}(u)$ & $\psi_{10}(u)$ & $\psi_{1}(u)$ & $\psi_{10}(u)$ & $\psi_{1}(u)$ & $\psi_{10}(u)$ & $\psi_{1}(u)$ & $\psi_{10}(u)$  \\
\midrule
10 & 0.204 & 0.403 & 0.198 & 0.325 & 0.171 & 0.371 & 0.143 & 0.266 \\
\midrule
11 & 0.258 & 0.541 & 0.272 & 0.516 & 0.21 & 0.497 & 0.202 & 0.442  \\
\midrule
12 & 0.315 & 0.687 & 0.354 & 0.721 & 0.259 & 0.636 & 0.27 & 0.649  \\
\midrule
13 & 0.379 & 0.818 & 0.44 & 0.878 & 0.307 & 0.77 & 0.343 & 0.829   \\
\midrule
15 & 0.51 & 0.964 & 0.606 & 0.991 & 0.417 & 0.943 & 0.5 & 0.982  \\
\midrule
17 & 0.635 & 0.997 & 0.746 & 0.9998 & 0.532 & 0.994 & 0.643 & 0.999  \\
\midrule
20 & 0.787 & 1 & 0.886 & 1 & 0.691 & 1 & 0.816 & 1  \\
\midrule
25 & 0.934 & 1 & 0.978 & 1 & 0.878 & 1 & 0.954 & 1   \\
\midrule
30 & 0.984 & 1 & 0.997 & 1 & 0.964 & 1 & 0.992 & 1  \\
\midrule
40 & 0.9996 & 1 & 1 & 1 & 0.998 & 1 & 1 & 1   \\
\midrule
50 & 1 & 1 & 1 & 1 & 1 & 1 & 1 & 1   \\
\midrule
\end{tabularx}
}
\caption{Underestimating $\lambda$ for 4 different cases, with $u=0$ and $\rho = 0.4$. 
\label{Sous-estimer rho avec u=0}}
\end{table}

As for the case where $\rho$ is underestimated, when the maximal value $\lambda_{max}$ is exceeded, the ruin probabilities increase quickly and reach 1 for a given threshold. Once again, when $\rho$ increases, the ruin probabilities decrease. 
What is more, if we had instead a comfortable initial reserve, namely $u=100$ as for the previous paragraph, then it would prevent from high risks if the underestimation of $\lambda$ is not too large.

However, we can remark that here, the process is in general more risky for fixed values of $\mu_X$ and $\mu_Y$ with two light-tailed distributions rather than with two heavy-tailed. Actually, the heavy tails are more risky when $\lambda$ is not too much underestimated, i.e. $\lambda \leq 13$.

\subsubsection{Conclusion on the underestimations}

We can draw the following conclusion by considering all the results found for $\psi_1(u)$ or $\psi_{10}(u)$ (which corresponds to a short or mid-term problem):
if we look at the results when $\rho$ or $\lambda$ increase (i.e., when $\chi$ decreases), then, until $\chi = - 20 \%$, it looks worse to underestimate $\rho$ when $\mu_X < \mu_Y$.

However, for all the other cases, it looks worse to underestimate $\lambda$ than $\rho$. One argument to justify this point is that, for the moment, we have only considered a value $\mu_Y$ which is of the same order of $\mu_X$, i.e. $0.5 \mu_X \leq \mu_Y \leq 2 \mu_X$. By taking $\mu_Y >> \mu_X$, or maybe by looking at a 20 or 30-period horizon, we might find the converse.

\subsection{Prediction for a ten-year horizon}

We consider here that a period corresponds to a year. 

In this section, depending on the pair of distributions we choose for $(X,Y)$, we try to identify the initial reserve $u$ which fits the best with having a one-year ruin probability $\psi_1(u) = 0.005$, and then we compute $\psi_{10}(u)$. 

Indeed, if an insurance company respects precisely the regulation proposed by Solvency II towards the one-year ruin probability, then one would like to know what the related risk is for the further years. Here, we look at a horizon of 10 years.

We keep $\lambda = \gamma_0 = 10$, and the same kind of pair of distributions as before. We fix $\rho = 0.1$. We choose a safety loading $\chi = 12.5\%$.

For Table \ref{tableau d'ingénierie inversée pour mu_X = 2 et mu_Y = 1}, we take $\mu_X = 2$, $\mu_Y = 1$, and $c=35$.

For Table \ref{tableau d'ingénierie inversée pour mu_X = 1 et mu_Y = 2}, we take $\mu_X = 1$, $\mu_Y = 2$, and we adapt $c$ to have the same safety loading: $c=36.25$.

\begin{table}[H]
\centering
\resizebox{0.7\textwidth}{!}{
\newcolumntype{C}{>{\centering\arraybackslash}X}
\begin{tabularx}{\textwidth}{|C|C|C|C|}
\toprule
Distribution & $u$  & $\psi_{1}(u)$ & $\psi_{10}(u)$ \\
\midrule
$X \sim \Par(2,2)$, $Y \sim \Par(3,2)$ & 93 & $0.00491 \quad $  $\pm \ 0.00057$ & $0.0527 \quad $  $\pm \, 0.001823$ \\
\midrule
$X \sim \Par(2,2)$, $Y \sim \Exp(1)$ & 88  & $0.00486 \quad $  $\pm \ 0.00057$ & $0.05353 \quad $  $\pm \, 0.00184$ \\
\midrule
$X \sim \Exp(1/2)$, $Y \sim \Par(3,2)$ & 37 & $0.00489 \quad $  $\pm \ 0.00057$ &  $0.09348 \quad $  $\pm \, 0.002375$ \\
\midrule
$X \sim \Exp(1/2)$, $Y \sim \Exp(1)$ & 33 & $0.00479 \quad $  $\pm \ 0.00056$ & $0.0931 \quad \ $  $\pm \, 0.0023$  \\
\midrule
\end{tabularx}
}
\caption{$\psi_{10}(u)$ if $\psi_1(u) \approx 0.005$, with $\chi = 12.5 \%$, $\mu_X =2$, and $\mu_Y = 1$. 
\label{tableau d'ingénierie inversée pour mu_X = 2 et mu_Y = 1}}
\end{table}

\begin{table}[H] 
\centering
\resizebox{0.7\textwidth}{!}{
\newcolumntype{C}{>{\centering\arraybackslash}X}
\begin{tabularx}{\textwidth}{|C|C|C|C|}
\toprule
Distribution & $u$  & $\psi_{1}(u)$ & $\psi_{10}(u)$ \\
\midrule
$X \sim \Par(3,2)$, $Y \sim \Par(2,2)$ & 93 & $0.00485 \quad $  $\pm \ 0.00057$ & $0.05866 \quad $  $\pm \, 0.001917$ \\
\midrule
$X \sim \Par(3,2)$, $Y \sim \Exp(1/2)$ & 37  & $0.005 \quad $  $\pm \ 0.000575$ & $0.09779 \quad $  $\pm \, 0.00242$ \\
\midrule
$X \sim \Exp(1)$, $Y \sim \Par(2,2)$ & 92 & $0.00496 \quad $  $\pm \ 0.00057$ &  $0.05668 \quad $  $\pm \, 0.001886$ \\
\midrule
$X \sim \Exp(1)$, $Y \sim \Exp(1/2)$ & 33 & $0.00469 \quad $  $\pm \ 0.00056$ & $0.0983 \quad \ $  $\pm \, 0.00243$  \\
\midrule
\end{tabularx}
}
\caption{$\psi_{10}(u)$ if $\psi_1(u) \approx 0.005$, with $\chi = 12.5 \%$, $\mu_X =1$, and $\mu_Y = 2$. 
\label{tableau d'ingénierie inversée pour mu_X = 1 et mu_Y = 2}}
\end{table}

In the examples of Table \ref{tableau d'ingénierie inversée pour mu_X = 2 et mu_Y = 1}, when $\mu_X > \mu_Y$, we can explain that $\psi_{10}(u)$ is greater when $X$ is light-tailed by the fact that solving the equation $\psi_1(u) = 0.005$ leads to a smaller value of $u$ (than when $X$ is heavy-tailed), and it may ensue after that $\psi_{10}(u)$ is greater for an exponential distribution because of this smaller value of $u$.

We then exchange the distributions of $X$ and $Y$ for each case in Table \ref{tableau d'ingénierie inversée pour mu_X = 1 et mu_Y = 2} to see what happens. Now, $\mu_Y > \mu_X$, and we remark that the results are almost the same, just by turning "$X$" into "$Y$".

\section{Changes with a pseudo INAR(2) equation}
\label{section faux INAR(2)}

Actually, the assumption made in Section \ref{section de présentation du modèle initial} that each main claim necessarily engenders a by-claim through the term $N_k$ in the expression of $M_k$ might seem "rough".

Indeed, in a general framework of insurance companies, it is not likely that every main claim would cause later losses. For the particular case of the PFAS issue, regarding the approach of Section \ref{section de motivation sur les PFAS}, we could say that not all the events of pollution by these compounds should lead to claims related with the cleaning of the environment or health care. 
For instance, it may refer to a case where the company agreed to insure the civil liability of the client but not the damages caused, because of some possible exclusions in the contract, or depending on the precise situation faced.
\medskip

Hence, we could refine the original model from Equation \ref{modèle classique en 1D avec BC et INAR(1)}, and modify the sequence $(M_k)_{k \in \N}$ expressed thanks to a process INAR(1), and write instead for $k \geq 1$:
\begin{equation}
\label{équation du faux INAR(2)}
M_k \ = \ \rho_1 \circ M_{k-1} \ + \ \rho_2 \circ N_k \ ,
\end{equation}
where $\rho_1, \rho_2 \in [0,1]$ ($\rho_1$ takes the role of $\rho$ before, we change the notation to avoid confusion). We may call it the \textit{"pseudo INAR(2) equation"}, since there are two thinning operations, but it does not correspond precisely to an INAR(2) process.

Thus, with a probability $\rho_2$, each main claim would now produce a second loss (or several subsequent claims gathered in a by-claim as an aggregate amount, according to Section \ref{section de présentation du modèle initial}).
We set $\rho_2 \neq 0$, so that the source term
of $M_k$ is not null.

We assume this is the only substitution we make toward the initial framework presented in Section \ref{section de présentation du modèle initial}.
\bigskip

In the remainder of this section, we explain how the results of Sections \ref{section avec les premières propriétés} to \ref{section de résultats de ruine avec queue lourde} can be adapted.

As we will see, many properties and results are generalised easily by changing the writing properly, in particular by turning $\lambda$ into $\rho_2 \lambda$. 

Moreover, the first adaptation concerns the expression of $\gamma_k$ (in Proposition \ref{gamma_n avec le faux INAR(2)} below), and according to this new expression, some of the results do not need to be modified.

\subsection{First properties (section \ref{section avec les premières propriétés})}

First of all, Proposition \ref{gamma_n} becomes:

\begin{prop}[New distribution of $(M_k)_{k \in \N}$] ~\\
\label{gamma_n avec le faux INAR(2)}
For $k \in \N$, $M_k \sim \cP(\gamma_k)$, while for $k \geq 1$, $\ \gamma_k = \rho^{k} \, \left( \gamma_0 - \frac{\rho_2\lambda}{1 - \rho_1} \right) + \frac{\rho_2 \lambda}{1 - \rho}$. \\
The sequence $(\gamma_k)_{k \in \N}$ is still monotonic, such that $\ \underset{k \to \infty}{\lim} \gamma_k = \frac{\rho_2 \lambda}{1 - \rho_1} , \ $ and \\
$M_k \overset{\cL}{\longrightarrow} \xi$, where $\xi \sim \cP \left( \frac{\rho_2 \lambda}{1 - \rho_1} \right)$.
\end{prop}
\medskip

It follows that $\gamma_{max} = \max \left( \gamma_0, \frac{\rho_2 \lambda}{1 - \rho_1} \right)$, and $0 \ < \ \rho_{max} \, = \, 1 - \frac{\rho_2 \lambda \mu_Y}{c - \lambda \mu_X} \,\ < \ 1 \ .$ \\
In particular, $\underset{\rho_2 \to 0^+}{\lim} \rhm \, = \, 1^-$.

Yet, the net profit condition is still expressed $ \ c > \lambda \,  \mu_X + \mu_Y \, \gamma_{max} \ .$
\bigskip

\newpage
As for the covariances, some of the changes are only due to the fact that $\rho$ is rewritten as $\rho_1$. Here there are:
\begin{itemize}
    \item For $n \geq k \geq 0$, $\cov(M_n, M_k) = \rho_1^{n-k} \gamma_k$.
    \item For $n \geq k \geq 1$, $\cov(M_n, N_k) = \rho_1^{n-k} \rho_2 \lambda$.
    \item $\cov(A_i,B_j) = \lambda  \mu_X   \mu_Y \rho_1^{j-i} \rho_2 \quad \text{for} \ 1 \leq i \leq j$.
    \item $\cov(B_i,B_j) = (\mu_Y)^2  \gamma_i \,  \rho_1^{j-i} \quad \text{for} \ 1 \leq i < j$
    \item The rest remains unchanged.
\end{itemize}

\subsection{Light-tailed distributions and softened dependent models (sections \ref{section ruin prob queues fines} and \ref{section de comparaison des modèles indépendants})}

We now assume that both $X$ and $Y$ have light-tailed distributions. \\
We first adapt Proposition \ref{tfdeLaplace/FGM en 1D}:

\begin{prop}[New m.g.f. for $R^D$] ~\\
Let $t,s \geq 0$ be real numbers, and let us assume that $\rho_1 M_Y(s) < 1$, and $M_X(t) < \infty$.
Thus, the moment-generating function becomes:
\begin{equation}
\begin{aligned}
\cM_{S_n}(t,s) \ &= \ \exp \left[ \gamma_0 [B_n \left( M_Y(s) \right) - 1] \frac{\,}{\,} \right.   \\
&+ \left. \, \lambda \left(M_X(t) \left(n(1-\rho_2) \,+ \, \rho_2 M_Y(s) \, \frac{1 - B_n \left( M_Y(s) \right) + n  (1 - \rho_1)}{1 - M_Y(s) \, \rho_1} \right) - n \right) \right] \ .
\end{aligned}
\end{equation}
where $B_0(s) = 1, \ \text{then} \ \ B_i(s) = \rho_1  s \, B_{i-1} (s) + (1 - \rho_1) \ $ for $i \geq 1$.
\end{prop}

Then, the associated Lundberg equation is
$$c \, t \ = \ \lambda \left[ M_X(t) \left(1 - \rho_2 + \rho_2 \, \frac{ M_Y(t) (1 - \rho_1)}{1 - \rho_1 M_Y(t) } \right) -1 \right] \ .$$

Let us now assume that $X \sim \operatorname{Exp}(\eta)$, and $Y \sim \operatorname{Exp}(\delta)$. Then, the adjustment coefficient of the dependent model (Equation \ref{kappa d'ajustement modèle original (dépendant) 1D}) becomes:
$$\kappa_D \ = \ \frac{\eta + \delta(1 - \rho_1)}{2} \ - \ \frac{\lambda + \sqrt{c^2 [\eta - \delta (1-\rho_1)]^2 + \lambda^2 + 2 \lambda c [\eta(2\rho_2 -1) + \delta (1 - \rho_1)]}}{2c} \ .$$
Accordingly, we could also adapt the lower bounds of $\kd$ from Proposition \ref{majoration kappa 1D}, by taking into account $\rho_2$, and of course all the bounds by rewriting $\rho$ in $\rho_1$.
\bigskip

We keep the same assumptions for the softened model $(R^I_n)_{n \in \N}$. In particular, \\ $\E(M_k^I) = \gamma_k$, with the new value of $\gamma_k$ taken in Proposition \ref{gamma_n avec le faux INAR(2)}.

Then, for $(R^S_n)_{n \in \N}$, it remains almost the same: we keep $\E(M_k^S) = \gamma_k$, but this time \\
$M_k^S \, = \, \rho_2 \circ N_k + O_k$, and $O_k \sim \cP(\gamma_k - \rho_2\lambda)$. Of course, we have the same independence assumptions between the sequences $(N_k)_{k \in \N}$ and $(O_k)_{k \in \N}$.

By using $\overline{\rho_1} = 1 -\rho_1$, and $\overline{\rho_2} = 1 -\rho_2$, we find the new expressions:
$$\ki \ = \ \frac{\eta + \delta }{2} \, - \, \frac{\lambda(\rho_2 + \overline{\rho_1})+\sqrt{4 \lambda^2 \rho_2 \overline{\rho_1} + [\lambda (\rho_2 - \overline{\rho_1}) + c \overline{\rho_1} (\eta - \delta)]^2}}{2c(1-\rho_1)} \ ,$$
$$\ks \ = \ \frac{\eta + \delta }{2} \, - \, \frac{\lambda (\rho_2 + \overline{\rho_1} .\overline{\rho_2}) + \sqrt{4c \lambda (\delta - \eta \overline{\rho_2})\overline{\rho_1}^2 + [\lambda(\rho_2 + \overline{\rho_1} . \overline{\rho_2}) + c \overline{\rho_1} (\eta - \delta)]^2}}{2c(1-\rho_1)} \ .$$

In particular, we can check that these new expressions of the coefficients $\kd$, $\ki$ and $\ks$ fit with the previous when $\rho_2 = 1$.

Regarding the variances, we just have to switch $\rho$ into $\rho_1$, and $\lambda$ into $\rho_2 \lambda$ (except for the term $n \lambda \mu_{(X^2)}$ in each case), and thus we adapt the expressions and the differences calculated.

Regarding the comparison of the coefficients, the framework established in Section \ref{section de comparaison des modèles indépendants} (Proposition \ref{comparaison des 3 coefficients d'ajustement}) is still true:

\begin{prop}[Comparison of the 3 coefficients] ~\\
For a fixed $\rho_2 \in ]0,1]$, and for $\rho_1 \in [0, \rhm[$:
$$\kd(0,\rho_2) = \ks(0,\rho_2) < \ki(0,\rho_2) \quad \text{and} \quad \kd(\rho_1,\rho_2) < \ks(\rho_1,\rho_2) < \ki(\rho_1,\rho_2) \ .$$
\end{prop}

\begin{proof}
To underline the reliance of the models on the corresponding parameters, we write $R_n^D(\rho_1,\rho_2,\lambda)$, $R_n^S(\rho_1,\rho_2,\lambda)$ or $R_n^I(\rho_1,\rho_2,\lambda)$.

We have the equality in distribution, for each $n \in \N$:
$$ R^D_n (\rho_1, \, \rho_2, \, \lambda) \ \ =_{\cL} \ \ R^D_n (\rho_1, \, 1, \, \rho_2\lambda) \ - \ \sum\limits_{k=1}^n \sum\limits_{i=1}^{L_k} Z_{k,i} \ .$$
We have taken two sequences $(Z_{k,i})_{(k,i) \in \N^2}$ and $(L_k)_{k \in \N^*}$ which are i.i.d., mutually independent, and independent of the other sequences $X$, $Y$, $N$ and $M^D$. Moreover, $Z_{1,1} \sim X_{1,1}$, and \\ $L_1 \sim \cP((1 - \rho_2)\lambda)$.

We could write the same thing for $R^S$ or $R^I$, using the same compound sum, and with the same type of dependence assumptions.

Since $\kd (\rho_1, \, 1,\,  \rho_2 \lambda) \leq \ks (\rho_1, \, 1,\,  \rho_2 \lambda) \leq \ki (\rho_1, \, 1,\,  \rho_2 \lambda)$ thanks to Proposition \ref{comparaison des 3 coefficients d'ajustement}, we get the same framework as in this proposition and we can check that, for a fixed $\rho_2 \in ]0,1]$, the equality stands only between $\kd$ and $\ks$ when $\rho_1 = 0$.
\end{proof}

\subsection{Heavy-tailed distributions (section \ref{section de résultats de ruine avec queue lourde})}

The main adjustments concern Corollaries \ref{ruine ultime discrète avec 2 lois sous-exp} and \ref{1D Pareto ruine ultime}.

\begin{cor}[New results for subexponential asympt. ruin. prob.] ~\\
We assume that $X$ is such that its stationary excess distribution $X_e$ is subexponential, and \\
$\overline{G}(x) = O(\overline{F}(x))$, or conversely. \\
Then, when $u \to \infty$, the ultimate ruin probability satisfies:
\begin{equation}
\psi_I(u) \ \sim \ D_1  \int_u^\infty \left(\overline{F}(x) + \overline{G}\left( \frac{(1-\rho_1)x}{\rho_2}\right) \right)  dx  , \ \, \text{where} \ \, D_1 \ = \ \frac{\lambda(1-\rho_1)}{c(1-\rho_1) - \lambda[(1-\rho_1)\mu_X + \rho_2\mu_Y]} \  .
\end{equation}
\end{cor}

\begin{proof}
The proof of Corollary \ref{ruine ultime discrète avec 2 lois sous-exp} may be generalized with no great difficulty.

The main adaptations are the fact that $L_1 \sim \cP\left( \frac{\rho_2 \lambda}{1 - \rho_1} \right)$, and the process $(P_t)_{t \in \R_+}$ becomes $\ P_t \ = \ u \, + \, c\,t \, - \, \sum\limits_{i=1}^{M(t)} \left( X_i + \frac{\rho_2}{1-\rho_1}Y_i \right) \ \ \text{for} \ t \in \R_+ \ .$
\end{proof}

\newpage
\begin{cor}[New equivalents with Pareto distributions]
Keeping the assumptions of Corollary \ref{1D Pareto ruine ultime}, we get
$$D_1 \ = \ \frac{\lambda  (1 - \rho_1) (\alpha -1) (\beta -1) }{c(1-\rho_1)(\alpha-1)(\beta -1) - \lambda[(1-\rho_1)\theta(\beta -1) + \rho_2 \eta(\alpha -1)]} \ .$$

i) If $\alpha < \beta$, therefore $\psi_I(u) \ \sim \ D_1 \, \frac{\theta^\alpha}{\alpha - 1} \, u^{-\alpha + 1} \ $.
\smallskip

ii) If $\alpha > \beta$, then $\psi_I(u) \ \sim \ D_1 \, \frac{\eta^\beta \rho_2^\beta }{(1-\rho_1)^\beta(\beta -1)} \, u^{-\beta + 1} \ $.
\smallskip

iii) If $\alpha = \beta$, then
$\psi_I(u) \ \sim \ D_1 \, \frac{\theta^\alpha  (1 - \rho_1)^\alpha + \eta^\alpha \rho_2^\alpha}{(1 - \rho_1)^\alpha (\alpha - 1)} \, u^{-\alpha + 1} \ .$
\end{cor}

The other changes are:
\begin{itemize}
    \item In the proof of Proposition \ref{ERV en temps infini}, $\lambda$ has to be replaced by $\rho_2 \lambda$, and the same thing for the proof or the results from Proposition \ref{inégalité proba ruine sous-exp modèle 1D classique} and Corollary \ref{corollaire Pareto ruine ultime modèle classique 1D}.
    \item In Equation \ref{équation d'encadrement BRV temps fini}, the lower bound becomes $H_{1,T}(u) \, \sim \, T \rho_2 \lambda \, \nu(A) \, \overline{F_0}(u) \ $.
    \item In Equation \ref{équation d'encadrement BRV temps infini}, the lower bound becomes $H_1(u) \, \sim \, \frac{\rho_2 \lambda \, \nu(A) L(u)}{[c - \rho_2 \lambda (\mu_X + \mu_Y)](\alpha - 1)} \, u^{-\alpha + 1} \, $.
\end{itemize}

The rest is unmodified.

\section{Uncertainty model}
\label{section d'ouverture sur le modèle d'incertitude}

As we said before, in the model of Equation \ref{modèle classique en 1D avec BC et INAR(1)}, we originally outlined that the term $N_k$ involved in the expression of $M_k$ stands for the new health or environmental damages due to the emissions of perfluorinated compounds occurring at period $k \geq 1$.
In the previous section, we proposed a modification of the source term, which could actually be written $\rho_2 \circ N_k$.
\medskip

Actually, regarding the specific case of the PFAS issue, we can have another point of view.
Indeed, it is clear that not all the emissions would generate some damages towards health or the environment, which would be identified or detected (not too late) and potentially covered. We particularly refer to the fact mentioned in Section \ref{section de motivation sur les PFAS} that, at the moment, some companies propose specific PFAS coverages only for rare cases, or even for no case at all.

Thus, we could see instead $N_k$ as the number of clients of a given company responsible for a polluting activity, and for who this activity is indeed detected and insured starting from the $k^{th}$ period. $M_k$ would accordingly correspond to the total series of treatments insured regarding health or the environment at period $k$.
\bigskip

Thus, in the following model, we introduce $N_k^\xi$ and $M_k^\xi$, which, in the PFAS context, are at period $k$ the "real" numbers of emissions and active contaminations engendered by the clients of a given company. In addition, we apply a kind of filtration thanks to Bernoulli variables, to underline which damages have been actually observed and have triggered covered claims.
\smallskip

The parameter of uncertainty corresponds to a real number $0 < \xi < 1$. The new model is defined like this:
\begin{equation}
    R^\xi_n \ = \ u \ + \ c \, n \ \, - \ \, \sum\limits_{k=1}^n \sum\limits_{i=1}^{N^\xi_k} \beta_{k,i}  X_{k,i} \ \, - \ \, \sum\limits_{k=1}^n \sum\limits_{i=1}^{M^\xi_k} \beta_{k,i} Y_{k,i} \qquad \text{for} \ n \in \N \ ,
\end{equation}
The sequences $(\beta_{k,i})_{(k,i) \in \N^2}$ and $(N^\xi_k)_{k \in \N}$ are i.i.d., with $N_1^\xi \sim \cP \left( \frac{\lambda}{\xi} \right)$ and $\beta_{1,1} \sim \cB(\xi)$.
Moreover, we keep the INAR(1) formulation for the counting of the by-claims: \\ $M_{k+1}^\xi = \rho \circ M_k^\xi + N_k^\xi \,$ for $k \geq 1$, but this time with $M_0^\xi \sim \cP \left(\frac{\gamma_0}{\xi} \right)$.

Similarly to the original model, we assume that $X$, $Y$, and $\beta$ are mutually independent, but also that $N^\xi$ on the one hand, and $M^\xi$ on the other hand are independent of these 3 sequences.
\bigskip

Let us compare the two processes $(R_n)_{n \in \N}$ from Equation \ref{modèle classique en 1D avec BC et INAR(1)}, and $(R^\xi_n)_{n \in \N}$ from this section.

Let us note $A_k^\xi$ and $B_k^\xi$ the aggregate claim amounts respectively related to the main claims and the by-claims at period $k$ for $R^\xi$.

We find easily that $A_k$ and $A_k^\xi$ have the same distribution.

Then, concerning the by-claims, thanks to Proposition \ref{gamma_n}, we get $M_k \sim \cP(\gamma^\xi_k)$, for all $k \in \N$, where for $k \geq 1$, $\, \gamma_k = \rho^{k} \, \left( \frac{\gamma_0}{\xi} - \frac{\lambda}{\xi(1 - \rho)} \right) + \frac{\lambda}{\xi(1 - \rho)}$.

We have clearly $\gamma_k^\xi > \gamma_k$. Indeed, for each $k \in \N$, seen as a function of $\lambda$, $\lambda \mapsto \gamma_k(\lambda)$ is increasing. More precisely, $\gamma_k = \xi \,\gamma_k^\xi$.

Actually, due to the $(\beta_{k,i})_{(k,i) \in \N^2}$, we can write $B_k^\xi \sim C_k$, where $C_k = \sum\limits_{i=1}^{J_k^\xi} Y_{k,i}$, and the variables $(J^\xi_k)_{k \in \N}$ are independent, with $J_k^\xi \sim \cP(\xi \, \gamma_k^\xi)$.
All in all, $B_k \sim B_k^\xi$.

Therefore, $R_n \ =_{\cL} \ R_n^\xi$, hence the risks from both models are the same.
\bigskip

We could also compare with the idea of the pseudo INAR(2) process proposed in Section \ref{section faux INAR(2)}: if we apply Equation \ref{équation du faux INAR(2)} to $(R_n)_{n \in \N}$, with $\rho_1 = \rho$, $\rho_2 = \xi$, and $N_1 \sim \cP \left( \frac{\lambda}{\xi} \right)$ instead of $\cP(\lambda)$, it is still possible to draw a comparison.
We can prove that once again, $R_n \sim R_n^\xi$ for each $n \in \N$.
\bigskip

All in all, in the context of this section, $(R^\xi_n)_{n \in \N}$ has maybe a better concrete sense than $(R_n)_{n \in \N}$, however since the processes have the same distribution, all the results proved about the $R$ allows us to understand the behaviour of $R^\xi$, especially in terms of risks.
\smallskip

\newpage
\section{Conclusion and openings}

In this paper, we have written a new kind of discrete-time risk model in ruin theory, involving main claims and delayed by-claims counted thanks to a time series INAR(1), and this model could serve to represent the PFAS issue.

We acknowledge there are several things we have not taken into account. For instance, in the practical reality, the frequency $\lambda$, the coefficient $\rho$, and the distributions of the main claims and the by-claims should be dependent, but also depend on the time. Yet, we stated that the analysis should become (very) difficult as soon as we add even only one of such assumptions.

Otherwise, when the contagion is quite spread it could be interesting to focus on the "filiation" of damages engendered by an initial event of this phenomenon, in order to link through the mathematical notations a main claim with all the by-claims it has produced, or at least to count at each period the number of by-claims coming from each one of the previous period.

Besides, since these pollutants are called eternal and sometimes the effects take years to be detected, we could imagine replacing the INAR(1) process by an INAR(p) process, where $p \in \N^*$ could be a relatively great integer ($p \geq 20$ regarding the time scale of the issue).
Such an improvement would be another way to describe the idea of filiation.
Another justification for such a process is the fact that, in the databases of some insurance or reinsurance companies, some claims are registered with the year or the period during which the first damages occurred, rather than the time at which the problem was discovered. Yet, such modifications should not be simple to characterise if one aims at providing significant mathematical results.
On the other hand, one could consider that such a meaning is included in the INAR(1), since we could register the propagation of such a path through the time until the first period where it does generate financial losses, and meanwhile assigning correspondingly claims with a null value.

Furthermore, with reliable data in the future, one could check if this model is sufficiently accurate regarding the PFAS issue (possibly with a little bit of tuning), and if so, use it to predict the behaviour and the viability of a corresponding insurance company.

All these ideas are left for future research.

\section{Acknowledgments}

This work benefited from the support of the Research Chair ACTIONS under the aegis of the Risk Foundation, a joint initiative by BNP Paribas Cardif and by the Institut des Actuaires.

This work was also conducted within the Research Chair DIALog under the aegis of the Risk Foundation, a joint initiative by Université Paris-Dauphine PSL and CNP Assurances.

\appendix

\newpage


\end{document}